\documentclass[12pt,american,french,english,preprint,authoryear]{elsarticle}
\usepackage[T1]{fontenc}
\usepackage[latin9]{inputenc}
\usepackage{color}
\usepackage{calc}
\usepackage{units}
\usepackage{amsmath}
\usepackage{amssymb}
\usepackage{graphicx}
\usepackage{esint}

\makeatletter

\newcommand{\noun}[1]{\textsc{#1}}
\providecommand{\tabularnewline}{\\}

\usepackage[a4paper,top=2.3cm,bottom=3cm,left=3.6cm,right=3.6cm]{geometry}

\journal{Journal of the Mechanics and Physics of Solids }

\@ifundefined{showcaptionsetup}{}{%
 \PassOptionsToPackage{caption=false}{subfig}}
\usepackage{subfig}
\makeatother

\usepackage{babel}
\addto\extrasfrench{%
   
   \providecommand{\fg}{\ifdim\lastskip>\z@\unskip\fi~\frqq}
}

\begin{document}
\begin{frontmatter}

\title{Elastic consequences of a single plastic event: towards a realistic
account of structural disorder and shear wave propagation in models
of flowing amorphous solids}

\maketitle

\author[label1,label2]{Alexandre NICOLAS}
\ead{alexandre.nicolas@ujf-grenoble.fr}
\author[label1,label2]{Francesco PUOSI}
\author[label1,label2]{Hideyuki MIZUNO\fnref{labh}}
\author[label1,label2,label3]{Jean-Louis BARRAT}

\address[label1]{Univ. Grenoble Alpes, LIPhy, F-38000 Grenoble, France}
\address[label2]{CNRS, LIPhy, F-38000 Grenoble, France}
\address[label3]{Institut Laue-Langevin, 6 rue Jules Horowitz, BP 156, F-38042 Grenoble, France}
\fntext[labh]{Present address: German Aerospace Center (DLR), Institute of Materials Physics in Space, Linder Höhe, 51147 Köln}
\begin{abstract}
Shear transformations\emph{ }(\emph{i.e.}, localised rearrangements
of particles resulting in the shear deformation of a small region
of the sample) are the building blocks of mesoscale models for the
flow of disordered solids. In order to compute the time-dependent
response of the solid material to such a shear transformation, with
a proper account of elastic heterogeneity and shear wave propagation,
we propose and implement a very simple Finite-Element (FE) - based
method. Molecular Dynamics (MD) simulations of a binary Lennard-Jones
glass are used as a benchmark for comparison, and information about
the microscopic viscosity and the local elastic constants is directly
extracted from the MD system and used as input in FE. We find very
good agreement between FE and MD regarding the temporal evolution
of the \emph{disorder-averaged} displacement field induced by a shear
transformation, which turns out to coincide with the response of a
\emph{uniform} elastic medium. However, \emph{fluctuations} are relatively
large, and their magnitude is satisfactorily captured by the FE simulations
of an elastically heterogeneous system. Besides, accounting for elastic
anisotropy on the mesoscale is not crucial in this respect.

The proposed method thus paves the way for models of the rheology
of amorphous solids which are both computationally efficient and realistic,
in that structural disorder and inertial effects are accounted for.\end{abstract}
\end{frontmatter}

\section{Introduction}

\textcolor{black}{Glasses are macroscopically isotropic and homogeneous.
Microscopically, the absence (or elusiveness) of a clear structural
signature of the liquid-to-glass transition upon cooling may fallaciously
lead one to believe that the property of structural homogeneity holds
down to the microscale in glasses, as it does in liquids. In the past
decades, it has become clear that this idea is completely erroneous:
structural disorder is often regarded as a crucial aspect not only
of the glass transition, but also of the flow of soft or hard glassy
materials, and more generally amorphous solids, such as emulsions,
foams, dense gels, and granular matter. For instance, several theories
of the glass transition, including the Random First Order Theory \citep{Lubchenko2007,Berthier2011review}
and kinetically constrained models with facilitated dynamics \citep{Chandler2010},
put the focus on dynamical heterogeneities, that is, the coexistence
of regions with fast and slow (arrested) dynamics.}

\textcolor{black}{Heterogeneities are even more manifest when the
materials are forced to flow. Instead of a homogeneous deformation,
one observes localised bursts of particle rearrangements, called shear
transformations or plastic events, embedded in an essentially elastically
deforming medium \citep{Argon1979,Falk1998,Schall2007,Amon2012}.
These irreversibly rearranging regions coincide with {}``weak''
zones where the local elastic (shear) moduli vanish at the onset of
a plastic event \citep{Tsamados2009}. By simply looking at the instantaneous
(static) configuration of the system, computing its soft modes, and
observing where they concentrate, one can predict statistically (but
only to a }\textcolor{black}{\emph{limited}}\textcolor{black}{{} extent)
the position of future rearrangements \citep{Widmer-Cooper2008,Rottler2014}.
Not only does microscopic structural disorder play the leading role
in fixing where the rearrangements will occur, but it also affects
the way stress is redistributed in the medium during these plastic
events, }\textcolor{black}{\emph{i.e.}}\textcolor{black}{, the propagation
of the shear waves originating from the rearranging region.}

\textcolor{black}{This stress redistribution is generally described
as the solution of an Eshelby inclusion problem in a uniform linear
elastic medium \citep{Eshelby1957}, with an inclusion that is often
assumed pointwise in lattice-based rheological models, for convenience
\citep{Picard2004,Picard2005,Vandembroucq2011,Talamali2011,Lin2014Density,Martens2011,Martens2012,Nicolas2014u}.
The solution is given by an elastic propagator $\mathcal{G}$ with
a characteristic four-fold angular symmetry and an $r^{-2}$ spatial
decay in two dimensions, in line with experiments on, }\textcolor{black}{\emph{e.g.}}\textcolor{black}{,
dense emulsions \citep{Desmond2013} (also see \citet{Budrikis2013,Sandfeld2015}
for a discussion on this elastic propagator and its possible numerical
implementations). However, some of us very recently showed that such
description only holds }\textcolor{black}{\emph{on average}}\textcolor{black}{{}
\citep{Puosi2014}; if an individual plastic event is considered,
the description is unreliable, because the average response is blurred
by sample-to-sample fluctuations, presumably associated with the elastic
heterogeneity of the material. Moreover, this approach neglects inertial
effects by supposing instantaneous mechanical equilibration, or, in
other words, an infinite shear wave velocity, whereas the role of
inertia on the statistics of avalanche sizes has been numerically
evidenced \citep{Salerno2012,Salerno2013}. These two deficiencies,
possibly among others, undermined a recent endeavour of ours to reproduce
the spatio-temporal correlations in the flow of a disordered solid
with a coarse-grained model using the elastic propagator $\mathcal{G}$
\citep{Nicolas2014s}.}

\textcolor{black}{The objective of this contribution is to go beyond
the average, equilibrium-based description in terms of the elastic
propagator; we aim to devise and put to the test a }\textcolor{black}{\emph{minimal
framework}}\textcolor{black}{{} allowing to capture the fluctuations
in the response due to structural disorder, as well as the propagation
of the shear waves, in two dimensions (2D). To this end, we implement
a basic Finite Element (FE) code and use Molecular Dynamics (MD) simulations
of an athermal solid as a benchmark. In so doing, we show how the
microscopic data about, }\textcolor{black}{\emph{e.g.}}\textcolor{black}{,
the local elastic constants can be extracted from the MD system and
used as input in FE.}

\textcolor{black}{In Section 2, we present the MD simulation method
and we introduce our simplified FE algorithm. Section 3 is concerned
with the fitting of the mechanical parameters required by FE, in particular,
the calculation of the local elastic constants of the MD solid. Section
4 clarifies the protocol to trigger artificial shear transformations.
Finally, Sections 5, 6, and 7 describe the disorder-averaged elastic
response to this localised transformation, the fluctuations around
this average, and the response in a particular configuration of the
system, respectively.}

\section{Methods}

\subsection{Molecular Dynamics}

To probe the flow properties of amorphous solids, we resort to MD
simulations of a 2D amorphous system. More precisely, we simulate
a binary mixture of \emph{A} and \emph{B} particles, with $N_{A}=32500$
and $N_{B}=17500$, of respective diameters $\sigma_{AA}=1.0$ and
$\sigma_{BB}=0.88$, confined in a square box of dimensions $205\sigma_{AA}\times205\sigma_{AA}$,
with periodic boundary conditions. The system is at reduced density
1.2. The particles, of mass $m=1$, interact via a pairwise Lennard-Jones
potential, 
\[
V_{\alpha\beta}\left(r\right)=4\epsilon_{\alpha\beta}\left[\left(\frac{\sigma_{\alpha\beta}}{r}\right)^{12}-\left(\frac{\sigma_{\alpha\beta}}{r}\right)^{6}\right],
\]
where $\alpha,\beta=A,\, B$, $\sigma_{AB}=0.8$,$\epsilon_{AA}=1.0$,
$\epsilon_{AB}=1.5$, and $\epsilon_{BB}=0.5$. The potential is truncated
at $r_{c}=2.5\sigma_{AA}$ and shifted for continuity.

We conduct our study in the athermal limit, by thermostatting the
system to zero temperature, so that no fluctuating force appears in
the equations of motion, \emph{viz.,} 
\begin{eqnarray}
\frac{d\boldsymbol{r_{i}}}{dt} & = & \boldsymbol{v_{i}}\nonumber \\
m\frac{d\boldsymbol{v_{i}}}{dt} & = & -\sum_{i\neq j}\frac{\partial V\left(r_{ij}\right)}{\partial\boldsymbol{r_{ij}}}+\boldsymbol{f_{i}}^{D}.\label{eq:eq_of_motion_MD}
\end{eqnarray}
\foreignlanguage{american}{The dissipative force $\boldsymbol{f_{i}}^{D}$
experienced by particle \emph{i} is computed with a Dissipative Particle
Dynamics (DPD) scheme, whereby particles are damped on the basis of
their relative velocities with respect to their neighbours. More precisely,
$\boldsymbol{f_{i}}^{D}$\emph{ }reads
\begin{eqnarray}
\boldsymbol{f_{i}}^{D} & = & -\sum_{j\neq i}\zeta w^{2}\left(r_{ij}\right)\frac{\boldsymbol{v_{ij}}\cdot\boldsymbol{r_{ij}}}{r_{ij}^{2}}\boldsymbol{r_{ij}}\label{eq:f_diss_DPD}\\
\text{where }w(r) & \equiv & \begin{cases}
1-\frac{r}{r_{c}} & \text{ if }r<r_{c},\\
0 & \text{ otherwise.}
\end{cases}\nonumber 
\end{eqnarray}
Here, $\boldsymbol{v_{ij}}\equiv\boldsymbol{v_{i}}-\boldsymbol{v_{j}}$
denotes the relative velocity of particle $i$ with respect to $j$,
the vector $\boldsymbol{r_{ij}}\equiv\boldsymbol{r_{i}}-\boldsymbol{r_{j}}$
connects particle \emph{j }to \emph{i}, the cut-off distance is set
to $r_{c}=2.5\sigma_{AA}$, and $\zeta$ controls the damping intensity.
Different values of $\zeta$ will be tested to probe the different
damping regimes, from underdamped ($\zeta\lesssim1$) to highly overdamped
($\zeta\gg1$). Note that, in Eq.~\ref{eq:f_diss_DPD}, the projection
of the force onto the radial vector $\boldsymbol{r_{ij}}$ is required
in order to conserve angular momentum. Several other virtues of DPD
have been exposed by \citet{Soddemann2003}. As far as we are concerned,
one of the main advantages is that, in the light of the recent work
of \citet{Varnik2014}, experimentally measured correlations in the
flow of amorphous solids are better reproduced numerically when dissipation
is based on \emph{relative} particle velocities, in opposition to
a mean-field damping scheme, in which \emph{absolute} velocities (with
respect to a hypothetic solvent flow) are used.} The impact of this
implementation on the propagation of shear waves will be discussed
in Section \ref{sub:Theoretical-expectations_prop}.

However, the DPD algorithm does not conserve the position of the centre
of mass of the system \emph{a priori}. Since the ensuing global translations
of the system may disturb the forthcoming analysis of displacements
in reponse to shear transformations, the system is regularly re-centred
during the simulation.

Equations \ref{eq:eq_of_motion_MD} are integrated with the velocity
Verlet algorithm with $\delta t=0.005$. In all the following, we
use $\tau_{LJ}\equiv\sqrt{m\sigma_{AA}^{2}/\epsilon}$ as the unit
of time and $\sigma_{AA}$ as the unit of length.

\subsection{Simplified Finite Elements\label{sec:FE_presentation}}

In the presence of elastic heterogeneities, the elastic response to
a localised shear transformation becomes intractable to analytical
calculations. This notably implies that the Fast Fourier Transform
routine commonly used in elastoplastic models needs to be replaced.
As a minimal substitute, we propose a simplified FE algorithm, which
will also allow us to account for inertial effects.

The FE method consists in discretising a Continuum Mechanics equation
onto a mesh. Here, the Continuum Mechanics equation involves elastic
and dissipative (viscous) forces, as well as inertia; hence, the momentum
conservation equation reads

\begin{equation}
\underset{\text{inertial force}}{\underbrace{\rho\frac{D\boldsymbol{\dot{u}}}{Dt}(\boldsymbol{r,}t)}}=\underset{\text{elasticity}}{\underbrace{\nabla\cdot\left[{\bf C}(\boldsymbol{r,}t)\boldsymbol{\epsilon}(\boldsymbol{r,}t)\right]}}+\underset{\text{viscosity}}{\underbrace{\eta\nabla^{2}\boldsymbol{\dot{u}}(\boldsymbol{r,}t)}},\label{eq:Continuum}
\end{equation}
where $\boldsymbol{u}$ and $\boldsymbol{\epsilon}$ are the displacement
and strain fields, respectively, $\nicefrac{D\bullet}{Dt}\equiv\nicefrac{\partial\bullet}{\partial t}+\left(\boldsymbol{v}\cdot\nabla\right)\bullet$
denotes the convected derivative, dots denote time derivatives, $\rho$
is the (area) density of the material, ${\bf C}$ denotes a local
stiffness matrix, and $\eta$ is the microscopic viscosity. Upon discretisation,
it turns into

\begin{equation}
\underset{\text{inertial force}}{\underbrace{\boldsymbol{\mathcal{M}}\cdot\ddot{u}}}=\underset{\text{elasticity}}{\underbrace{\boldsymbol{\mathcal{K}}\cdot u}}+\underset{\text{viscosity}}{\underbrace{\boldsymbol{\mathcal{H}}\cdot\dot{u}}},\label{eq:Discrete2}
\end{equation}
where $u$ is now a shorthand for the high-dimensional vector 
\[
\left(\begin{array}{ccccc}
u_{x}^{{\bf (N-1)}}, & u_{y}^{{\bf (N-1)}}, & \ldots & u_{x}^{{\bf (0)}}, & u_{y}^{{\bf (0)}}\end{array}\right)^{\top}
\]
containing the displacements along $x$ and $y$ at the $N$ nodes
of the mesh. $\boldsymbol{\mathcal{M}}$, $\boldsymbol{\mathcal{K}}$,
and $\boldsymbol{\mathcal{H}}$ are $2N\times2N$ real matrices (to
be specified later), and the dependences on time have been omitted.

\selectlanguage{american}%
\begin{figure}
\begin{centering}
\includegraphics[width=7cm]{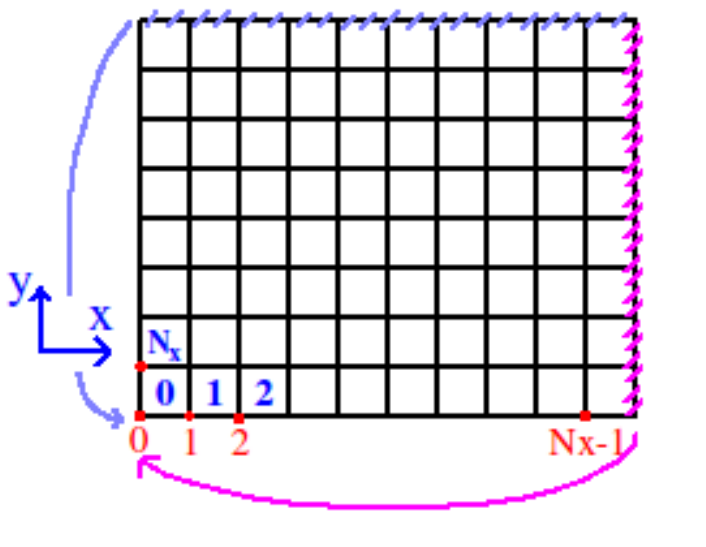}
\par\end{centering}

\caption{\label{fig:mesh_FEM}Sketch of the FE mesh. The system is periodic
in both directions, so that column $N_{x}$ coincides with column
0 and row $N_{y}$ coincides with row 0. There are $N=N_{x}\times N_{y}$
nodes and elements.}
\end{figure}

\selectlanguage{english}%
Bearing in mind our pursuit of minimalism, we choose a simple (static)
regular square meshgrid, as sketched in Fig.~\ref{fig:mesh_FEM}.
In an element, the local strain $\boldsymbol{\epsilon}\equiv\left(\begin{array}{ccc}
\epsilon_{xx}, & \epsilon_{yy}, & \sqrt{2}\epsilon_{xy}\end{array}\right)^{\top}$, using condensed notations for 2D symmetric tensors, is a function
of the displacements at the local nodes, and we make the approximation
of a \emph{uniform} strain within each element%
\footnote{\selectlanguage{french}%
In practice, our simplified FE method is therefore close to a Finite
Volume method.\selectlanguage{english}
}. For convenience, let us number these nodes from 0 to 3 counter-clockwise,
for a given element, starting from the bottom left corner, \emph{viz.},
\foreignlanguage{french}{${3\atop 0}{\scriptstyle \square}{2\atop 1}$}.
In an analogous way, the (uniform) elemental stress $\boldsymbol{\sigma}^{\mathrm{el}}$
is derived from the nodal forces $\left(f_{x}^{\mathrm{el}},f_{y}^{\mathrm{el}}\right)$.
Since the mesh is regular, we can define a constant $3\times8$ real
matrix ${\bf B}$ that relates, in a given element, the (nodal) displacements
to the (elemental) strains, on the one hand, and the (nodal) forces
to the (elemental) stresses, on the other hand, \emph{viz.},

\begin{equation}
\left(\begin{array}{c}
\epsilon_{xx}\\
\epsilon_{yy}\\
\sqrt{2}\epsilon_{xy}
\end{array}\right)={\bf B}\cdot\left(\begin{array}{c}
u_{x}^{(0)}\\
u_{y}^{(0)}\\
\vdots\\
u_{x}^{(3)}\\
u_{y}^{(3)}
\end{array}\right)\text{ and }\left(\begin{array}{c}
\sigma_{xx}^{\mathrm{el}}\\
\sigma_{yy}^{\mathrm{el}}\\
\sqrt{2}\sigma_{xy}^{\mathrm{el}}
\end{array}\right)=-{\bf B}\cdot\left(\begin{array}{c}
f_{x}^{\mathrm{el}\,(0)}\\
f_{y}^{\mathrm{el}\,(0)}\\
\vdots\\
f_{x}^{\mathrm{el}\,(3)}\\
f_{y}^{\mathrm{el}\,(3)}
\end{array}\right).\label{eq:B_transp_matrix_FEM-1}
\end{equation}
The expression of the matrix ${\bf B}$ is given in \ref{app:FE_routine},
along with further details pertaining to the implementation of the
FE routine and the computation of the matrices \foreignlanguage{french}{$\boldsymbol{\mathcal{M}}$,
$\boldsymbol{\mathcal{K}}$, and $\boldsymbol{\mathcal{H}}$ appearing
in Eq.~\ref{eq:Discrete2}.} Note that the $\sqrt{2}$ prefactors
have been introduced with foresight (see Section \ref{sec:local_el_csts})
and the {}``minus'' sign preceding ${\bf B}$ in Eq.~\ref{eq:B_transp_matrix_FEM}
is due to the fact that $\boldsymbol{f}^{\mathrm{el}\,(i)}$ is the
force exerted \emph{by} the element \emph{on} node $i$.

\bigskip{}

The resulting routine is still simple enough to be used quite efficiently
in a coarse-grained model. In particular, (see \ref{app:discretisation_dynamics}),
the global force-displacement matrix is constant and, accordingly,
only has to be inverted \emph{once}, at the beginning of the simulation.

\bigskip{}
On the other hand, there are naturally a few downsides to this simplicity.
First and foremost, it is only marginally stable, insofar as the convergence
of the discrete FE solution to the continuous solution of Eq.~\ref{eq:Continuum}
is not guaranteed when the mesh size tends to zero. Consequently,
this scheme is not suited to general purpose. However, as will be
shown below, it is both satisfactory and very convenient for the modelling
of (the response to) shear transformations, where elements represent
material regions of finite size. In particular, the frequently encountered
checkerboard issue, whereby high and low displacements/velocities
alternate erratically in neighbouring cells (hence the image of a
checkerboard), is practically circumvented, provided that shear transformations
span four adjacent elements (a {}``macro-element'') and inertia
is present, \emph{i.e.}, $\rho\neq0$.

\section{Fitting of elastic and viscous parameters}

We are now left with the task of fitting the physical parameters appearing
in Eq.~\ref{eq:Continuum} with the MD parameters. Neglecting mesoscopic
density fluctuations, the density $\rho$ and the miscroscopic viscosity
$\eta$ are supposed to be constant, while the stiffness matrix ${\bf C}(\boldsymbol{r,}t)$
is allowed to vary in space.

\subsection{Viscosity}

To fit the viscosity $\eta$ in Eq.~\ref{eq:Continuum}, we compare
the stress due to homogeneous shear, at a rate $\dot{\gamma}$, as
calculated, on the one hand, in FE ($\sigma_{xy}=\eta\dot{\gamma}$),
and, on the other hand, in MD (where it is obtained through the Irving-Kirkwood
formula). The calculations are shown in their full extent in \ref{app:viscosity_fitting}
and lead to the following formula for a binary mixture of A and B
components: 
\[
\eta=\frac{\pi}{4}\zeta\int_{0}^{\infty}\left[n_{A}^{2}g_{AA}(r)+2n_{A}n_{B}g_{AB}(r)+n_{B}^{2}g_{BB}(r)\right]w^{2}\left(r\right)r^{3}dr,
\]
where $n_{A}$ and $n_{B}$ are the number densities of $A$ and $B$
constituents in the system, $g_{AA}$, $g_{BB}$, and $g_{AB}$ are
the radial distribution functions for the $A-A$, $B-B$, and $A-B$
correlations, respectively, and $\zeta$ and $w$ are the DPD damping
coefficient and the damping function defined in Eq.~\ref{eq:f_diss_DPD}.

For the MD system under consideration, we obtain
\[
\eta=0.726\,\zeta.
\]

\subsection{Local elastic constants\label{sec:local_el_csts}}

\selectlanguage{american}%
Having determined the dissipative coefficient of the model, we turn
our attention to the \emph{local} elastic properties of the system.

The only relevant material lengthscale in the model being the typical
size ($a=5\sigma_{AA}$) of a rearrangement \citep{Nicolas2014s},
we tile the system into subregions of size $a$ and compute the local
stiffness tensors on this {}``mesoscopic'' scale, with \foreignlanguage{english}{the
local stress-affine strain method presented in Ref.~\citep{Mizuno2013moduli}.
Details of this protocol and issues related to the rather unfamiliar
local stiffness tensors are discussed in \ref{app:local_stiffness_tensors}.
With condensed notations, these tensors can be written as $3\times3$
real matrices in 2D, \emph{viz.},
\begin{equation}
\left(\begin{array}{c}
\sigma_{xx}\\
\sigma_{yy}\\
\sqrt{2}\sigma_{xy}
\end{array}\right)=\underset{{\bf C}}{\underbrace{\left(\begin{array}{ccc}
C_{xx,xx} & C_{xx,yy} & C_{xx,xy}\\
C_{yy,xx} & C_{yy,yy} & C_{yy,xy}\\
C_{xy,xx} & C_{xy,yy} & C_{xy,xy}
\end{array}\right)}}\left(\begin{array}{c}
\epsilon_{xx}\\
\epsilon_{yy}\\
\sqrt{2}\epsilon_{xy}
\end{array}\right),\label{eq:C3x3_FEM}
\end{equation}
where $\sigma_{xx}$, $\sigma_{yy}$, and $\sigma_{xy}$ are the linear
elastic contributions to the local stress. }

\selectlanguage{english}%
Contrary to their macroscopic counterpart, the local ${\bf C}$ matrices
are not symmetric \emph{a priori}, for very small regions \citep{Tsamados2009}.
However, the coarse grain $a=5\sigma_{AA}$ is large enough here for
the assumption of symmetry to be a reasonable approximation. To limit
the number of parameters, we further assume that isotropic contraction/dilation
of the region only generates an isotropic stress, \emph{i.e.}, that
\begin{eqnarray*}
\left(\begin{array}{ccc}
\epsilon_{xx} & \epsilon_{yy} & \sqrt{2}\epsilon_{xy}\end{array}\right)^{\top} & = & \nicefrac{\sqrt{2}}{2}\left(\begin{array}{ccc}
1 & 1 & 0\end{array}\right)^{\top}
\end{eqnarray*}
is an eigenvector of ${\bf C}$.

These two assumptions, namely, tensorial symmetry and isotropy of
the response to contraction, imply that the stiffness tensor should
be of the form
\begin{eqnarray}
{\bf C} & = & \left(\begin{array}{ccc}
\alpha & \delta & \beta\\
\delta & \alpha & -\beta\\
\beta & -\beta & \upsilon
\end{array}\right),\label{eq:C_full_proj_FEM}
\end{eqnarray}
where the parameters $\alpha,\delta,\beta,\upsilon\in\mathbb{R}$
are assessed in \ref{app:local_stiffness_tensors}. By analogy with
the macroscopic situation, the eigenvalues $c_{1}\leqslant c_{2}\leqslant c_{3}$
of the approximated matrix ${\bf C}$ are related to the local shear
moduli $\mu_{1}$ and $\mu_{2}$ and the local bulk modulus $K$ \emph{via}
$c_{1}=2\mu_{1}$, $c_{2}=2\mu_{2}$, and $c_{3}=2K$, and there exists
a frame $\left(\boldsymbol{e_{x}}(\theta),\,\boldsymbol{e_{y}}(\theta)\right)$,
rotated by an angle $\theta$ with respect to the original frame,
in which the stiffness tensor reads
\[
\left(\begin{array}{ccc}
K+\mu_{2} & K-\mu_{2} & 0\\
K-\mu_{2} & K+\mu_{2} & 0\\
0 & 0 & 2\mu_{1}
\end{array}\right),\text{ with }\mu_{1}\leqslant\mu_{2}.
\]
Consequently, the following four local parameters suffice to determine
${\bf C}$ completely: $\theta$, $\mu_{1}$, $\mu_{2}$, and $K$.

\bigskip{}

Table~\ref{tab:Statistical-prop} summarises the main features of
the distributions of $\mu_{1}$, $\mu_{2}$, and $K$ measured in
the Lennard-Jones glass under consideration; $\theta$ is uniformly
distributed, in accordance with macroscopic isotropy. 

It is noteworthy that the local stiffness matrices exhibit significant
anisotropy, as indicated by the discrepancy between the mean value
of the shear modulus in the (locally) weaker direction, $\left\langle \mu_{1}\right\rangle =13.16$,
and its strong counterpart, $\left\langle \mu_{2}\right\rangle =24.46$. 

Some regions actually even display negative shear moduli $\mu_{1}$.
This is not unrealistic in the MD system, because these regions can
be stabilised by the surrounding medium, but in the following they
will be discarded, and arbitrarily set to zero, in the FE simulations,
where they cause instabilities. 

Lastly, the bulk modulus is much larger (by a factor of 5) than the
shear moduli, in line with expectations, and its relative standard
deviation (\emph{i.e.}, the ratio of the standard deviation and the
mean value) is by far smaller than it is for the shear moduli, which
means that, on a relative basis, the latter are more broadly distributed.
Consequently, we will henceforth always neglect spatial fluctuations
of the bulk modulus and set $K=99.9$. As for the distributions of
shear moduli, three types of systems will be considered in FE:

(i) a uniform system, with \foreignlanguage{american}{$\mu_{1}=\mu_{2}=18.8$}

(ii) a heterogeneous system made of isotropic blocks ({}``het. iso.''),
with \foreignlanguage{american}{$\mu_{1}=\mu_{2}=18.8\pm5.3,$ \emph{i.e.},
a normal distribution of shear moduli\emph{ $\mu_{1}=\mu_{2}$ }with
mean value 18.8 and standard deviation 5.3. (Remember that each block
is a macro-element made of four adjacent finite elements.)}

(iii) a heterogeneous system made of anisotropic blocks ({}``het.
aniso.''), with \foreignlanguage{american}{$\mu_{1}=13.16\pm7.2\text{ and }\mu_{2}=24.46\pm5.8$
and a uniform distribution of the angles $\theta$.}

\begin{table}
\begin{centering}
\begin{tabular}{|c|c|c|c|}
\hline 
\emph{Denomination} & \emph{Symbol} & \emph{Mean } & \emph{Std dev.}\tabularnewline
\hline 
\hline 
Shear modulus (weak direction) & $\mu_{1}$ & 13.16 & 7.2\tabularnewline
\hline 
Shear modulus (strong direction) & $\mu_{2}$ & 24.46 & 5.8\tabularnewline
\hline 
Average shear modulus  & $\mu\equiv\frac{\mu_{1}+\mu_{2}}{2}$ & 18.81 & 5.3\tabularnewline
\hline 
Bulk modulus & $K$ & 99.9 & 8.4\tabularnewline
\hline 
\end{tabular}
\par\end{centering}

\caption{\label{tab:Statistical-prop}Statistical properties of the elastic
constant distributions: mean values and standard deviations (std dev.).}
\end{table}

\bigskip{}

Through the simulation of plane shear waves, we have checked that
the transverse sound velocity measured in FE is consistent with that
measured in MD.

\section{Protocol for the artificially triggered shear transformations}

In this section, we describe the protocol to artificially trigger
ideal shear transformations.

\bigskip{}

\selectlanguage{american}%
In the MD system, following \citet{Puosi2014}, shear transformations
are artificially created by applying a pure shear strain $\epsilon_{xy}$
to a disk centred at $(x_{0},y_{0})$ and of diameter $a=5\sigma_{AA}$.
To do so, particles whose initial position $(x_{i},y_{i})$ belongs
to this region are moved to a new position $(x_{i}^{\prime},y_{i}^{\prime})$
at $t=0$, which satisfies

\[
\begin{cases}
x_{i}\rightarrow x_{i}^{\prime} & =x_{i}+\epsilon_{xy}\left(y_{i}-y_{0}\right)\\
y_{i}\rightarrow y_{i}^{\prime} & =y_{i}+\epsilon_{xy}\left(x_{i}-x_{0}\right).
\end{cases}
\]
Their positions are then frozen for the whole simulation. In order
to measure the elastic,\emph{ i.e.}, reversible, response of the medium,
$\epsilon_{xy}$ never exceeds a few percent strain. Clearly, all
(transient or permanent) dilational effects \citep{Schuh2007} potentially
accompanying shear transformations are here discarded.

\begin{figure}
\begin{centering}
\includegraphics[width=4cm]{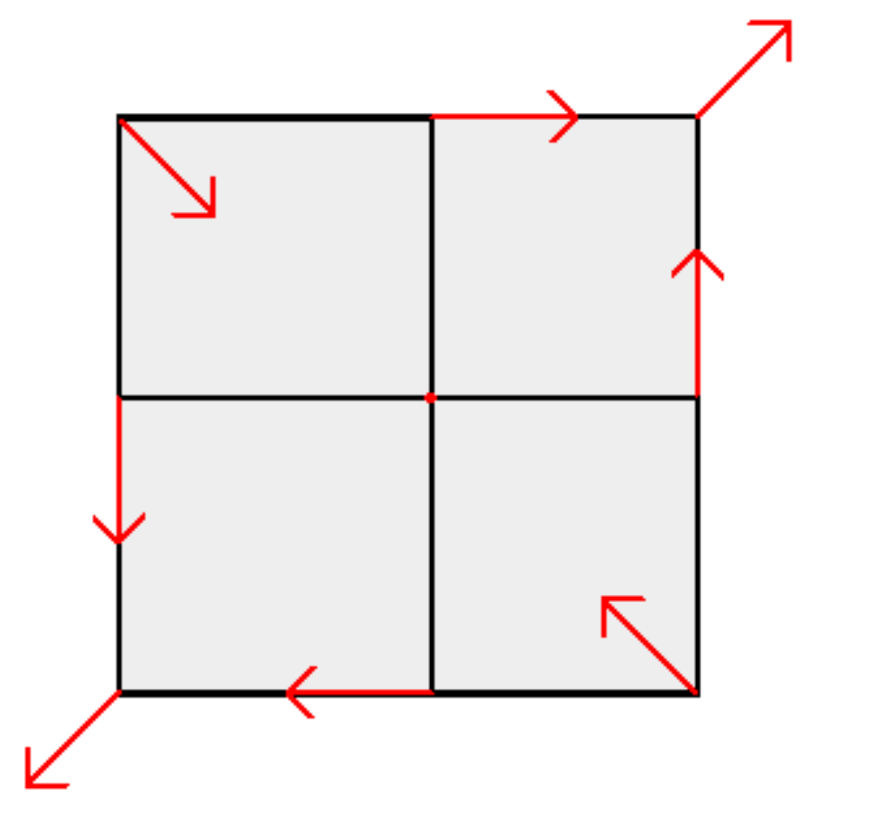}
\par\end{centering}

\caption{\label{fig:sketch_FE_STZ}Sketch of the displacements applied to a
macro-element to model a pure shear transformation.}
\end{figure}

A similar shear transformation is applied in the FE simulations to
a macro-element made of four adjacent elements (see Section~\ref{sec:FE_presentation}),
by controlling the positions of the nodes of these elements, as sketched
in Fig.~\ref{fig:sketch_FE_STZ}.

\selectlanguage{english}%

\section{Disorder-averaged propagation of shear waves\label{sec:Disorder-averaged-propagation}}

Let us first probe the \emph{disorder-averaged} time-dependent response
to a shear transformation, in different damping regimes, both in FE
and in MD. To this end, MD simulations are averaged over many (50)
locations of the shear transformation in the sample, while the FE
results are averaged over many (50) realisations of the disorder,
\emph{i.e., }of the random values of the local elastic constants.

\subsection{Comparison between MD and Finite Elements\label{sec:propagation_comp}}

\begin{figure}
\begin{centering}
\subfloat[$\zeta=1$ ($\eta=0.726$)]{\includegraphics[width=7cm]{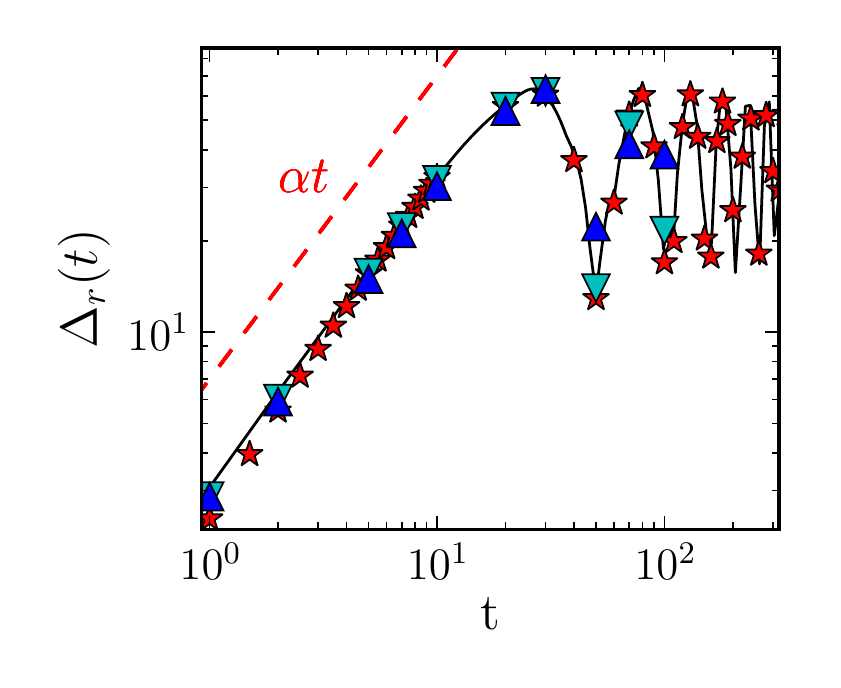}\includegraphics[width=7cm]{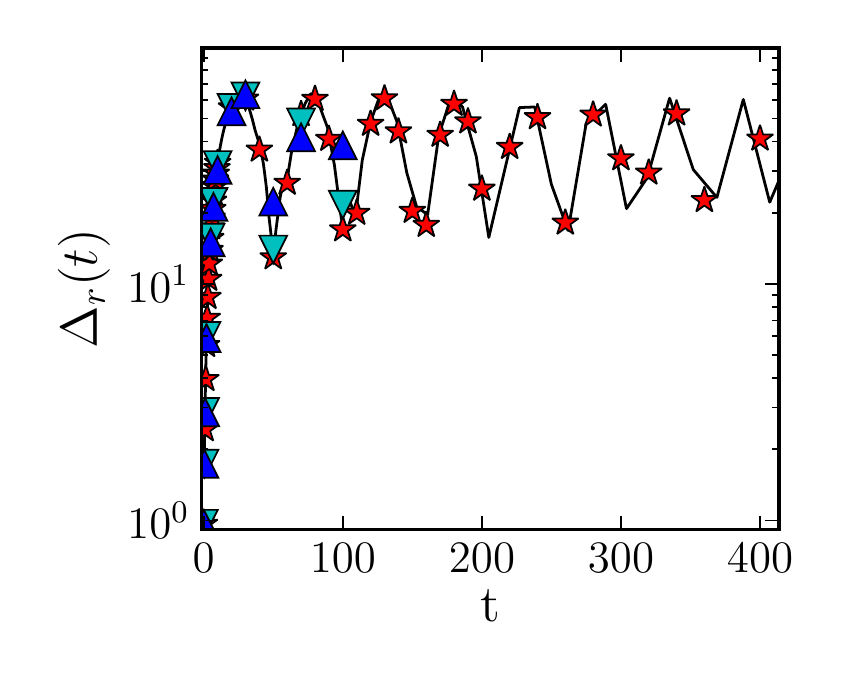}

}
\par\end{centering}

\begin{centering}
\subfloat[$\zeta=10$ ($\eta=7.26$)]{\includegraphics[width=7cm]{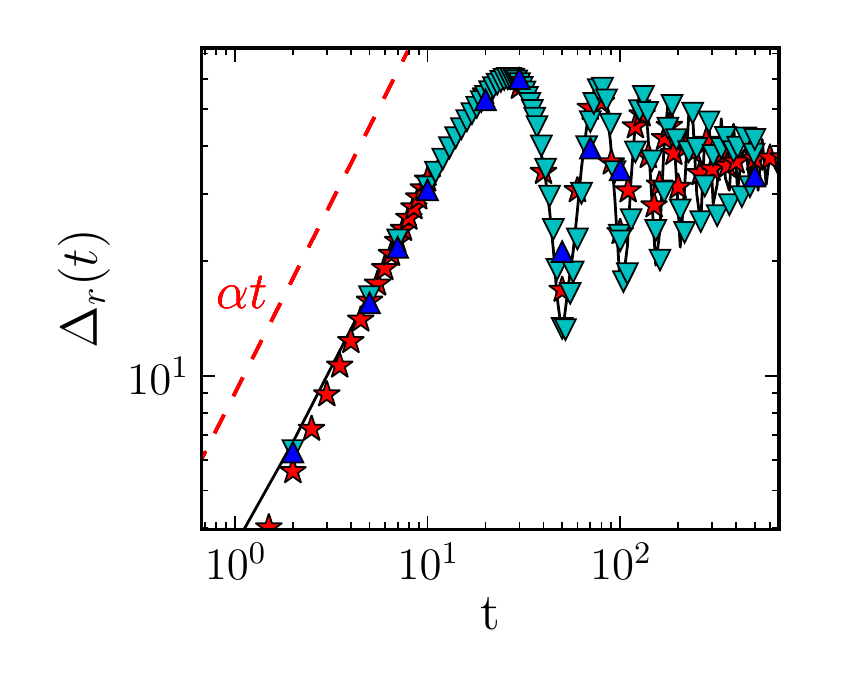}\includegraphics[width=7cm]{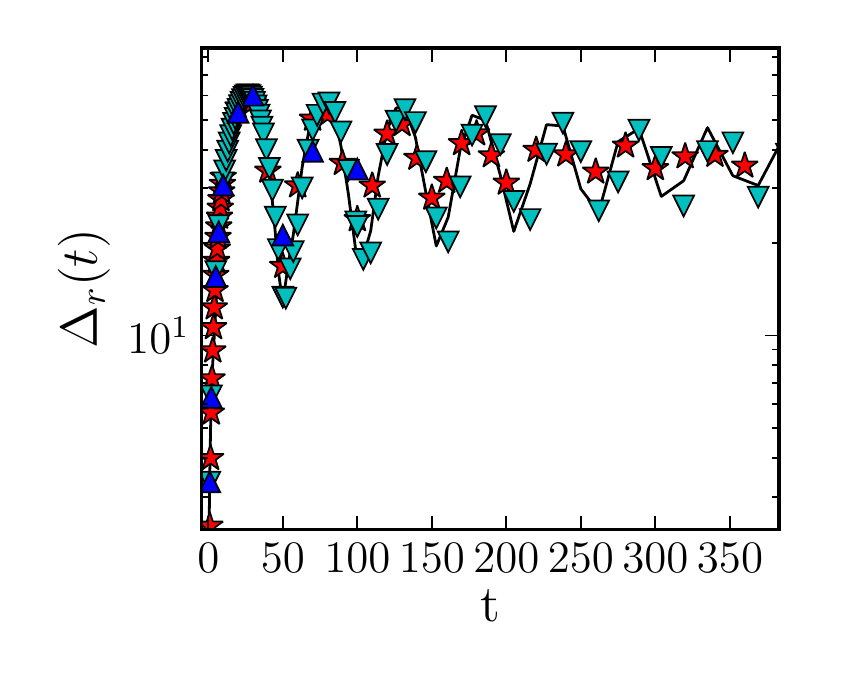}

}
\par\end{centering}

\begin{centering}
\subfloat[$\zeta=100$ ($\eta=72.6$)]{\includegraphics[width=7cm]{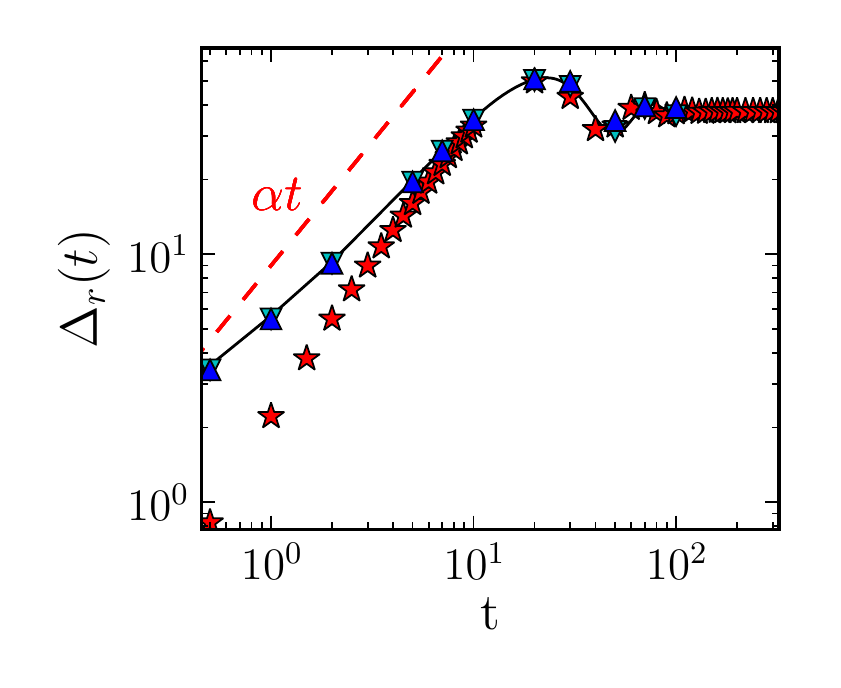}\includegraphics[width=7cm]{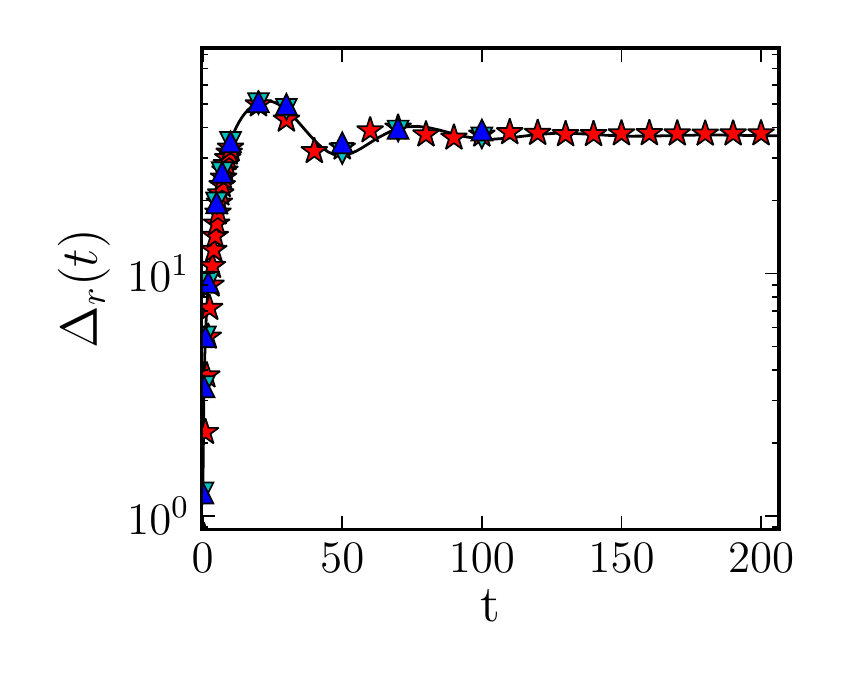}

}
\par\end{centering}

\caption{\label{fig:avg_prop_radius_FEM}Average propagation radius $\Delta_{r}$
as a function of time, for different damping magnitudes. \protect \\
(\emph{Red stars}) MD data; (\emph{inverted cyan triangles}) FE, het.
iso.; (\emph{blue triangles}) FE, het. aniso; (\emph{solid black line})
FE, uniform system. \protect \\
(\emph{Left}) log-log plot, (\emph{right}) same data, in semi-logarithmic
plot.}
\end{figure}

\selectlanguage{american}%
For a quantitative study, we make use of the average propagation radius
$\Delta_{r}(t)$ introduced by \citet{Puosi2014} to measure the advance
of the wave,
\[
\Delta_{r}(t)\equiv\iint|u_{r}(\boldsymbol{r};t)|d^{2}\boldsymbol{r},
\]
where $u_{r}(t)$ is the radial displacement at time $t$. If the
final displacement ($u_{r}(\boldsymbol{r};t=\infty)\sim r^{-1}$ in
any given direction $\theta$ in the far field) is essentially achieved
as soon as a region is reached by the wavefront, $\Delta_{r}(t)$
will grow linearly with the (linear) size of the displaced region.
The average propagation radius is plotted in Fig.~\ref{fig:avg_prop_radius_FEM}
for diverse values of the damping $\zeta$. The initial growth is
ballistic in MD, with $\Delta_{r}(t)\sim t$ , while at long times
$\Delta_{r}(t)$ saturates to its steady-state value. The evolution
of $\Delta_{r}(t)$ before the steady state is reached strongly depends
on $\zeta$. At low damping ($\zeta=1$), the interaction with the
waves generated by the periodic replicas of the shear transformation
leads to particularly long-lived oscillations of $\Delta_{r}(t)$
(Fig.~\ref{fig:avg_prop_radius_FEM}a), while stronger damping ($\zeta=100$)
completely suppresses these oscillations. 

The FE simulations nicely capture this qualitative change, and the
agreement both in the limit of low damping (Fig.~\ref{fig:avg_prop_radius_FEM}a)
and in the limit of strong damping (Fig.~\ref{fig:avg_prop_radius_FEM}c)
is excellent, at relatively long times. This is true for all three
FE systems, including the uniform one, which supports the idea that
the \emph{average} propagation in elastically heterogeneous media
is virtually identical to the propagation in a uniform medium. 

For an intermediate value of the damping, namely, $\zeta=10$ (Fig.~\ref{fig:avg_prop_radius_FEM}b),
the agreement is reasonable, but not quite as good, insofar as the
oscillations observed in MD are damped perceptibly faster than their
counterparts in FE, not only in the uniform system, but also in the
heterogeneous one (het. iso.). This suggests that the FE viscosity
is somewhat underestimated, or that the anharmonicities present in
MD significantly contribute to the damping of the oscillations. 

Finally, the short-time propagation is well described at low damping,
but the agreement declines when $\zeta$ increases, in which case
the FE method overestimates the propagation velocity over short distances.

\begin{figure}
\begin{centering}
\subfloat[$\Delta t=2$]{\includegraphics[width=6cm]{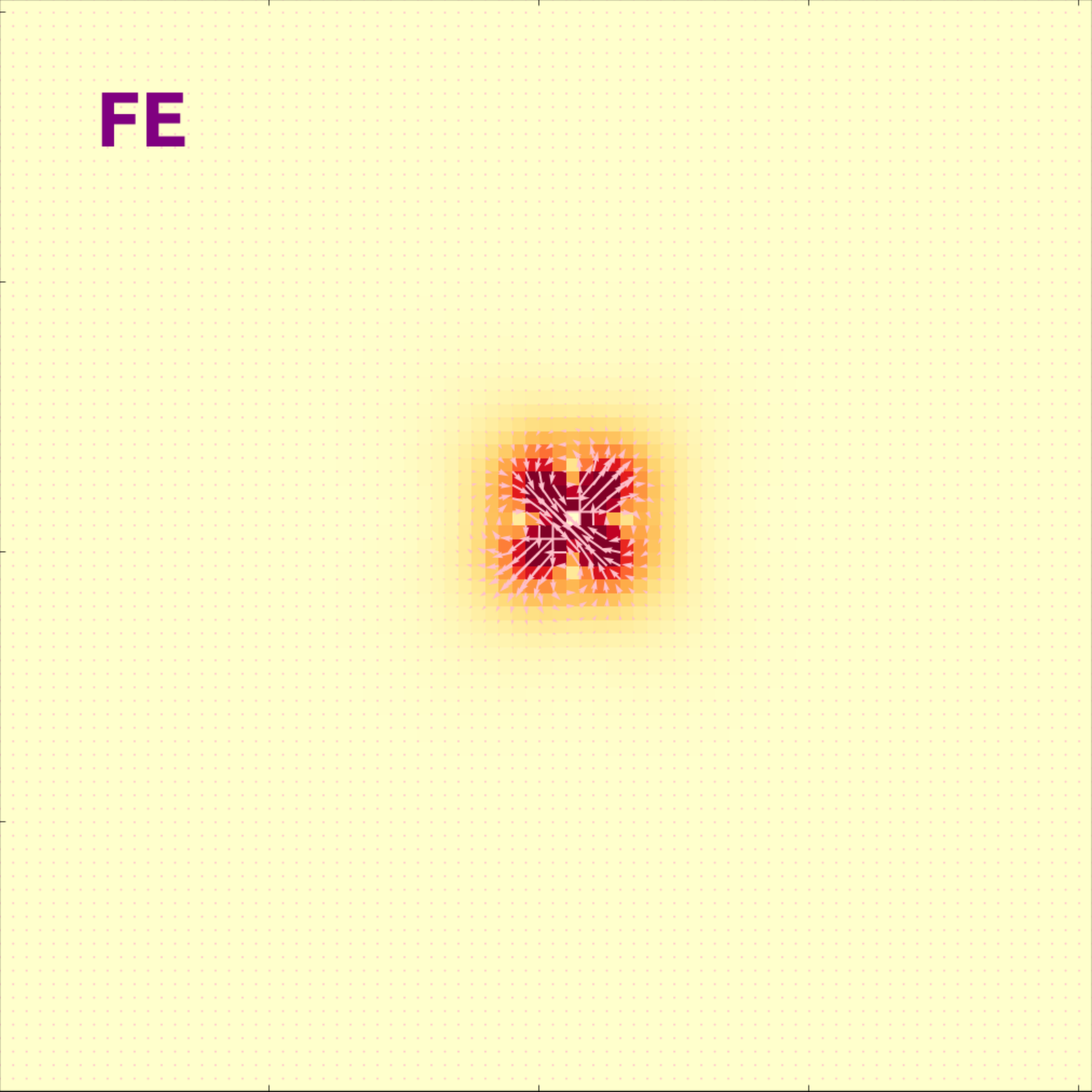}\hspace*{0.3cm}\includegraphics[width=6cm]{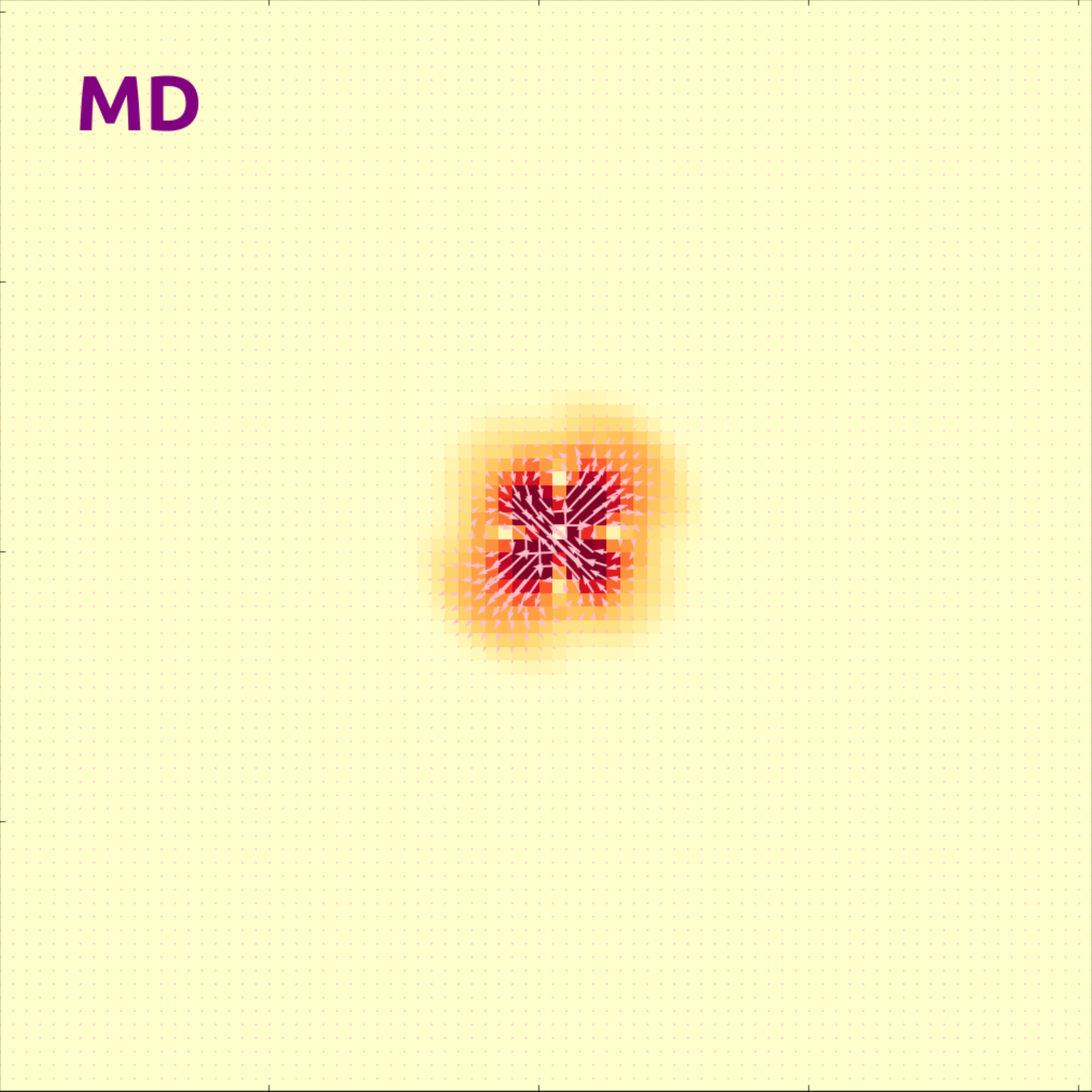}

}
\par\end{centering}

\begin{centering}
\subfloat[$\Delta t=10$]{\includegraphics[width=6cm]{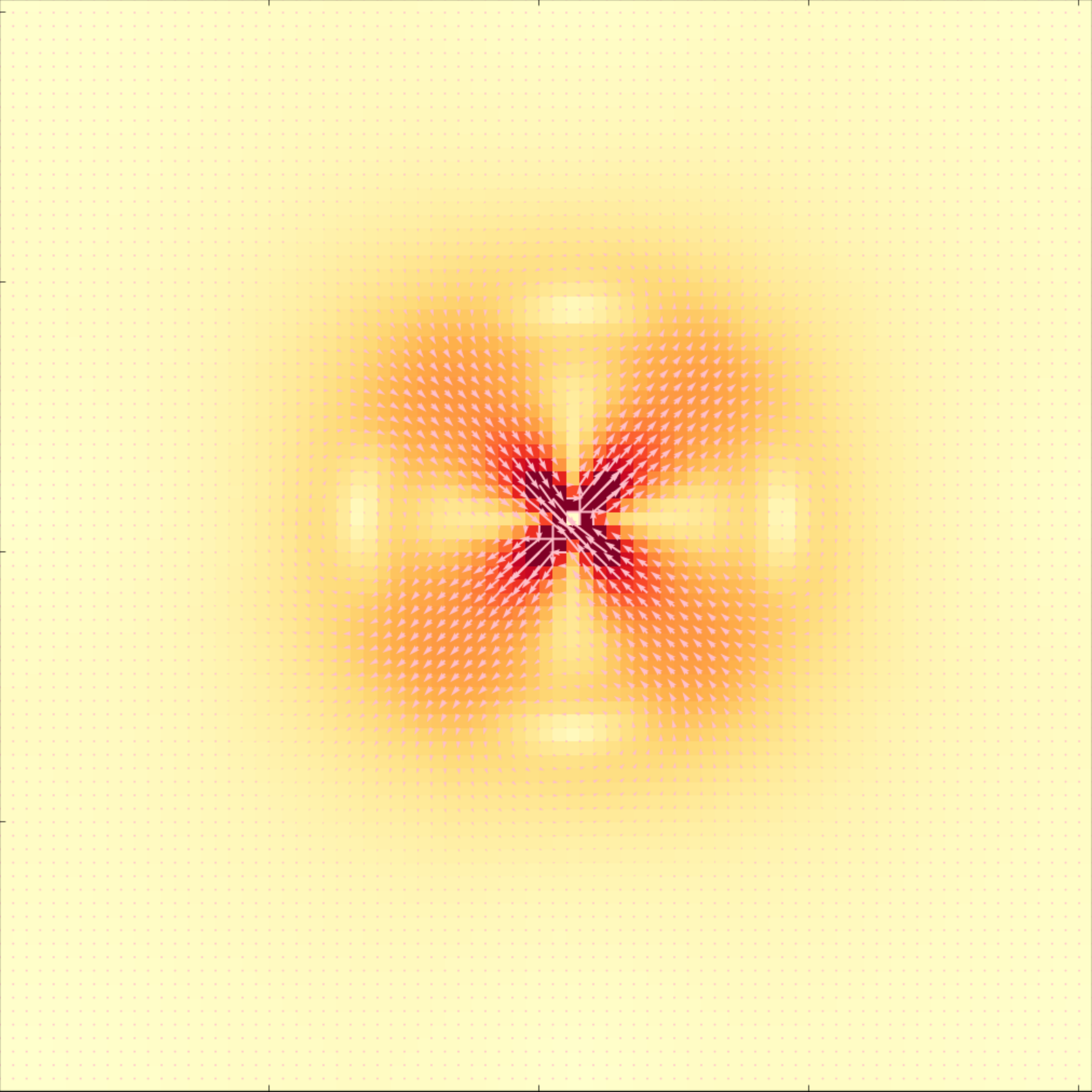}\hspace*{0.3cm}\includegraphics[width=6cm]{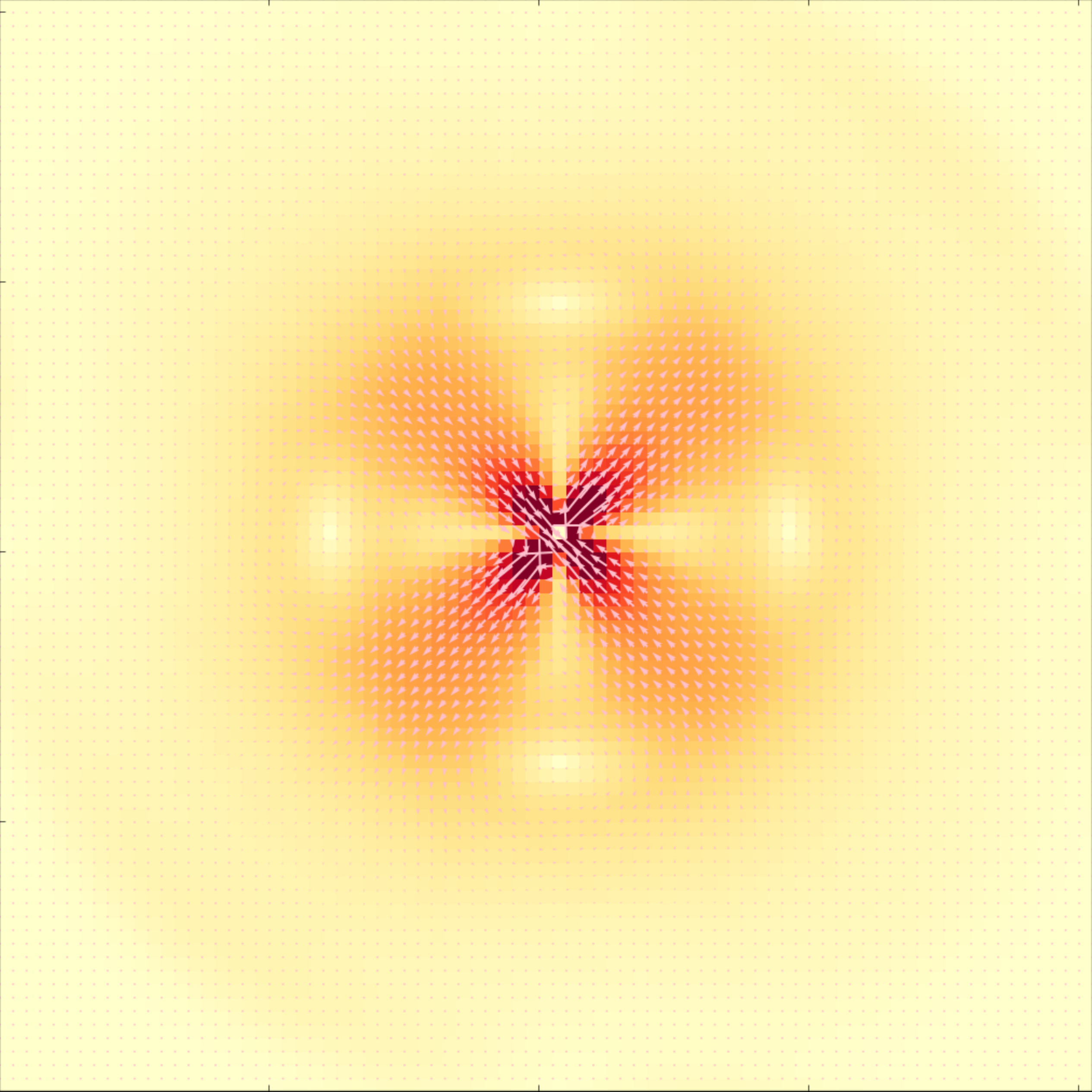}

}
\par\end{centering}

\begin{centering}
\subfloat[$\Delta t=1000$]{\includegraphics[width=6cm]{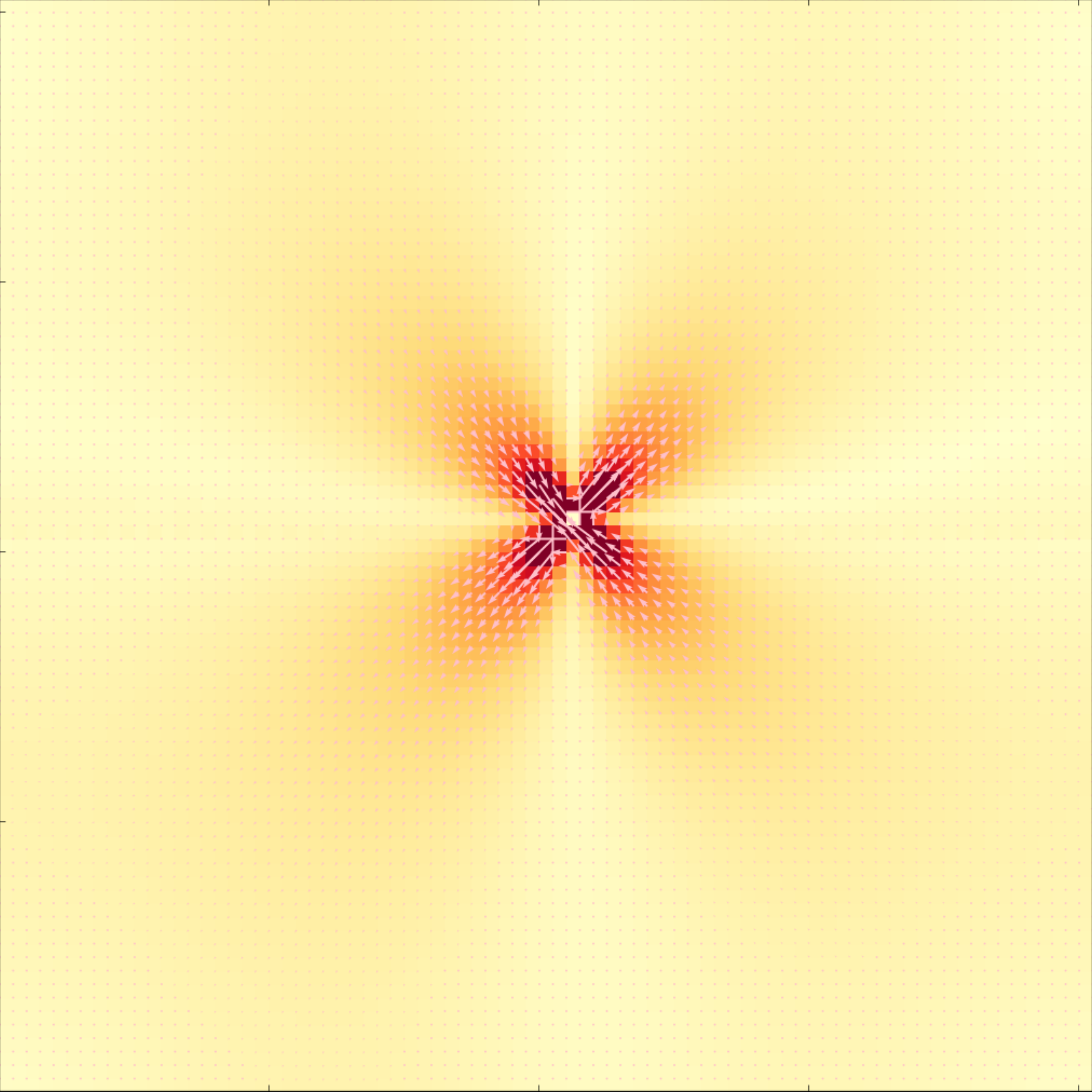}\hspace*{0.3cm}\includegraphics[width=6cm]{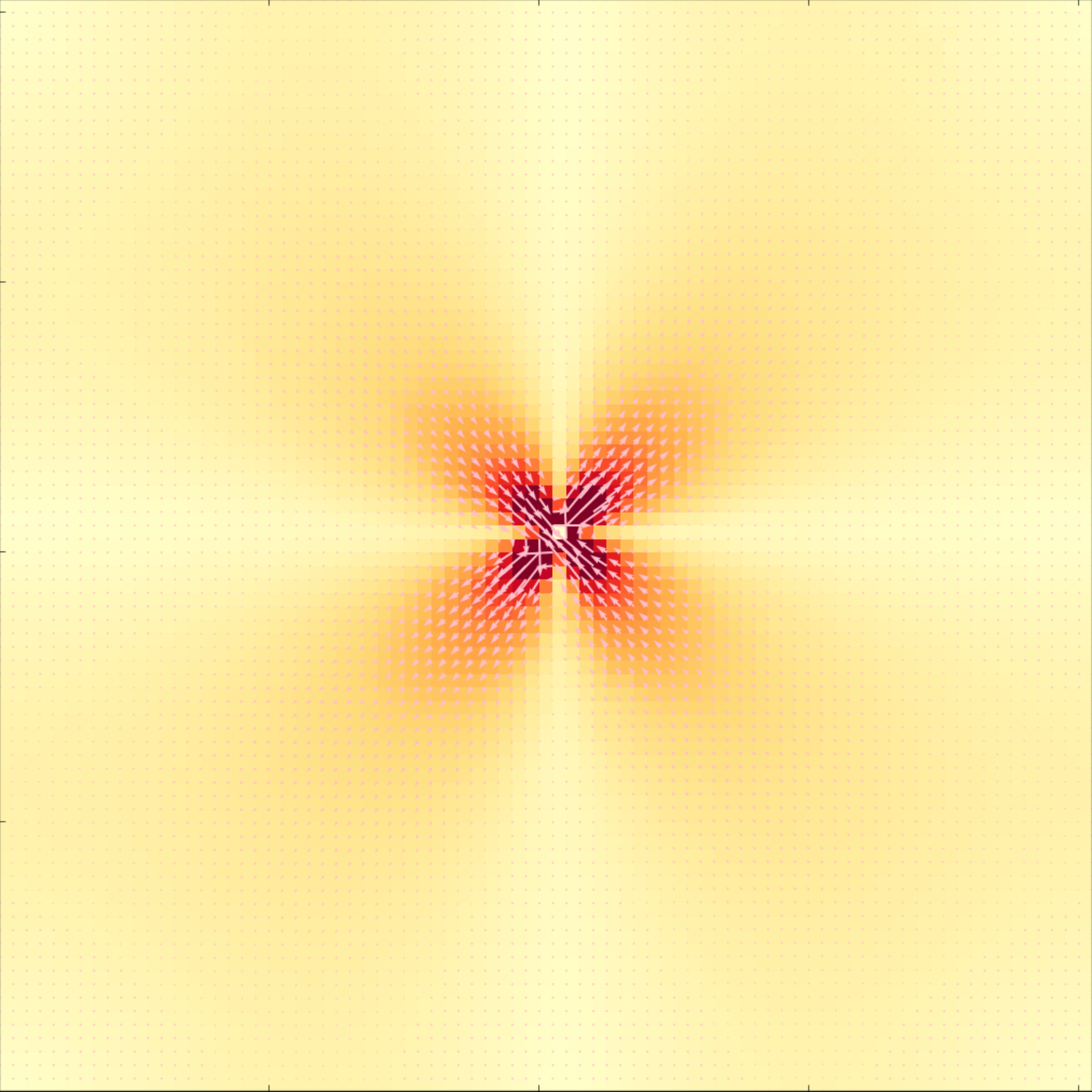}

}
\par\end{centering}

\caption{\label{fig:avg_disp_field_a_FEM}Average displacement field induced
by a shear transformation (at the centre of the cell), after a time
lag $\Delta t$, for relatively low damping, $\zeta=1$ (hence, $\eta=0.726$). The pink arrows represent the displacement
vectors and the background colour indicates their norms. System size:
$\left(205\sigma_{AA}\right)^{2}$, corresponding to $82^{2}$ finite
elements.\protect \\
(\emph{Left}) Finite Elements, het. iso.; (\emph{right}) Molecular
Dynamics. }
\end{figure}

\begin{figure}
\begin{centering}
\subfloat[$\Delta t=2$]{\includegraphics[width=6cm]{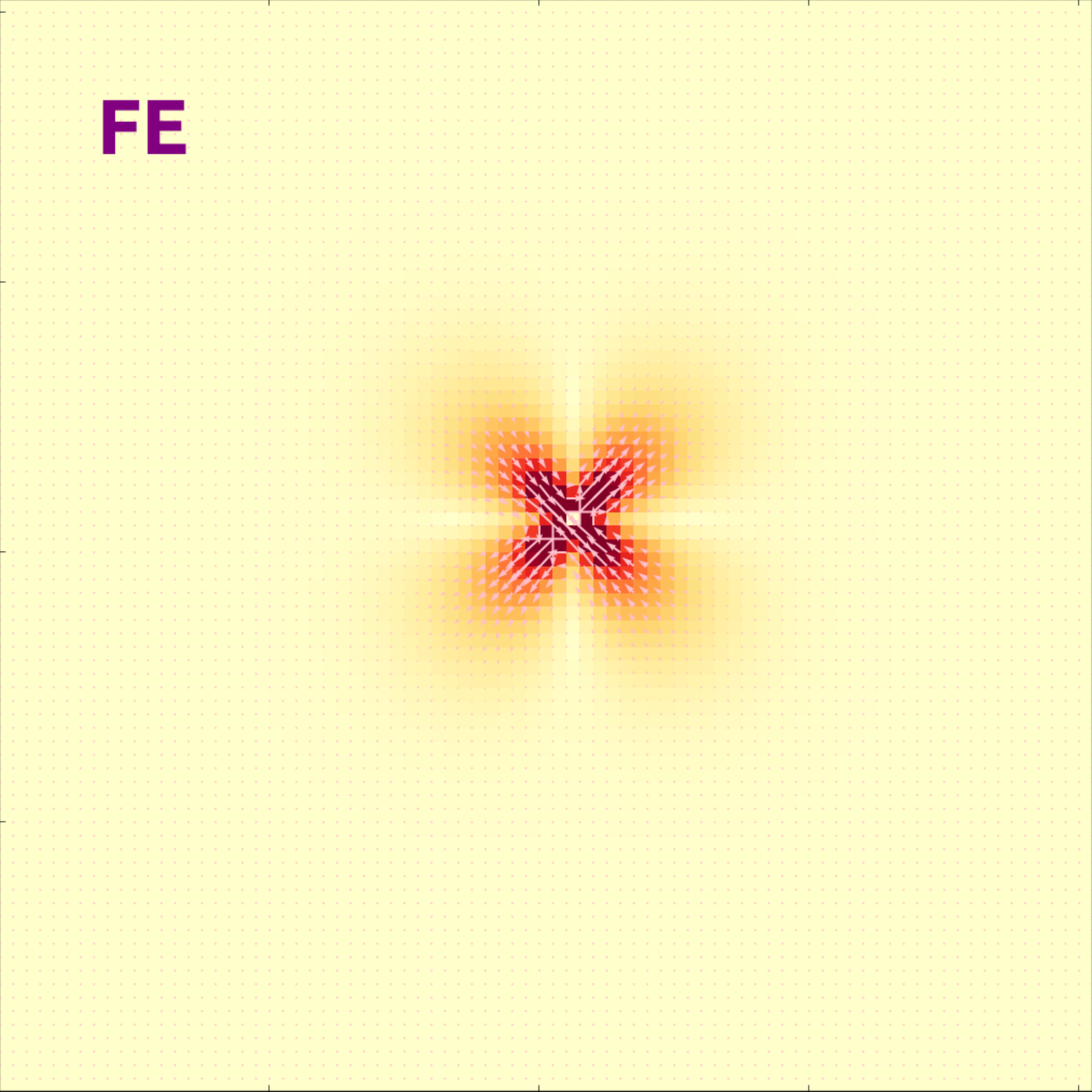}\hspace*{0.3cm}\includegraphics[width=6cm]{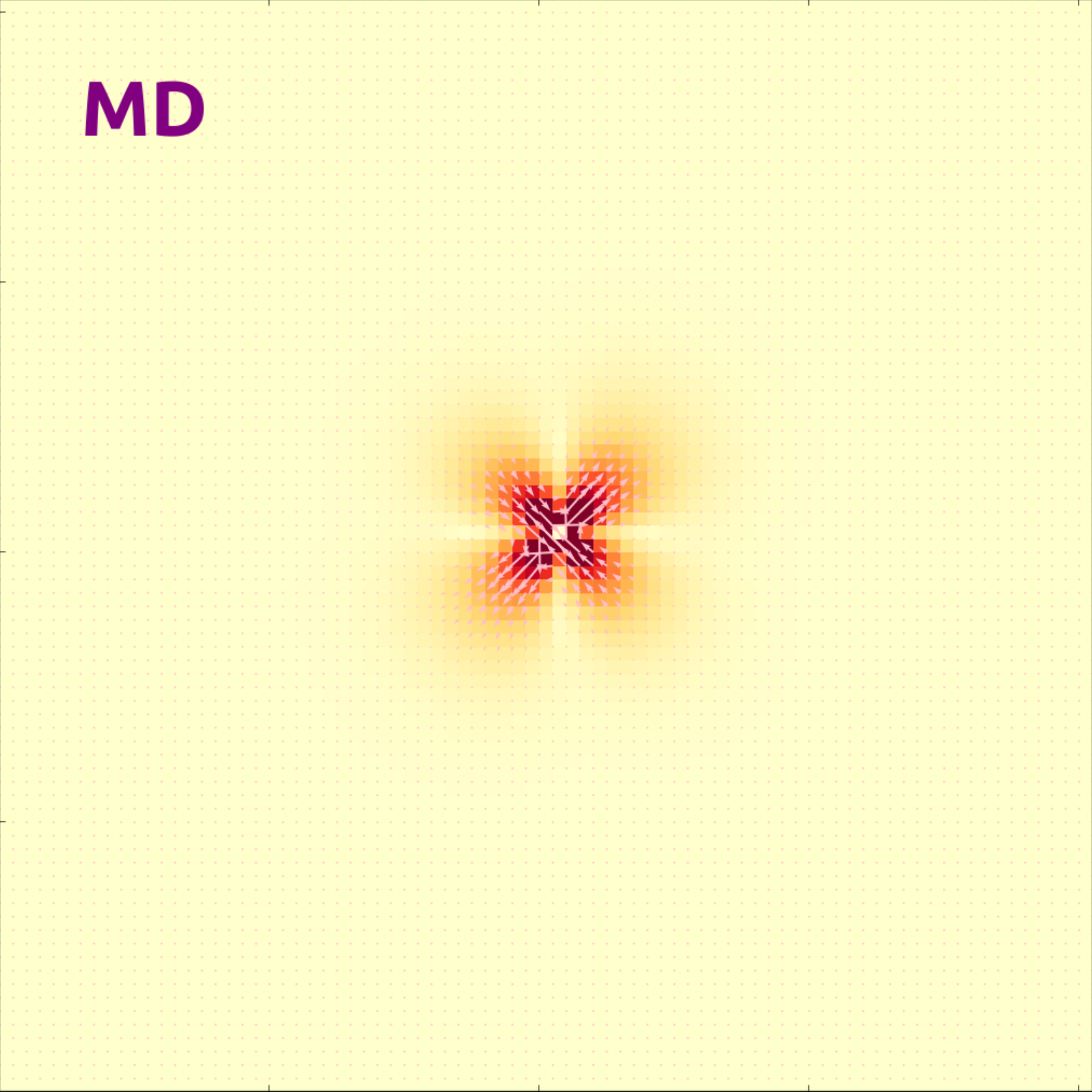}

}
\par\end{centering}

\begin{centering}
\subfloat[$\Delta t=10$]{\includegraphics[width=6cm]{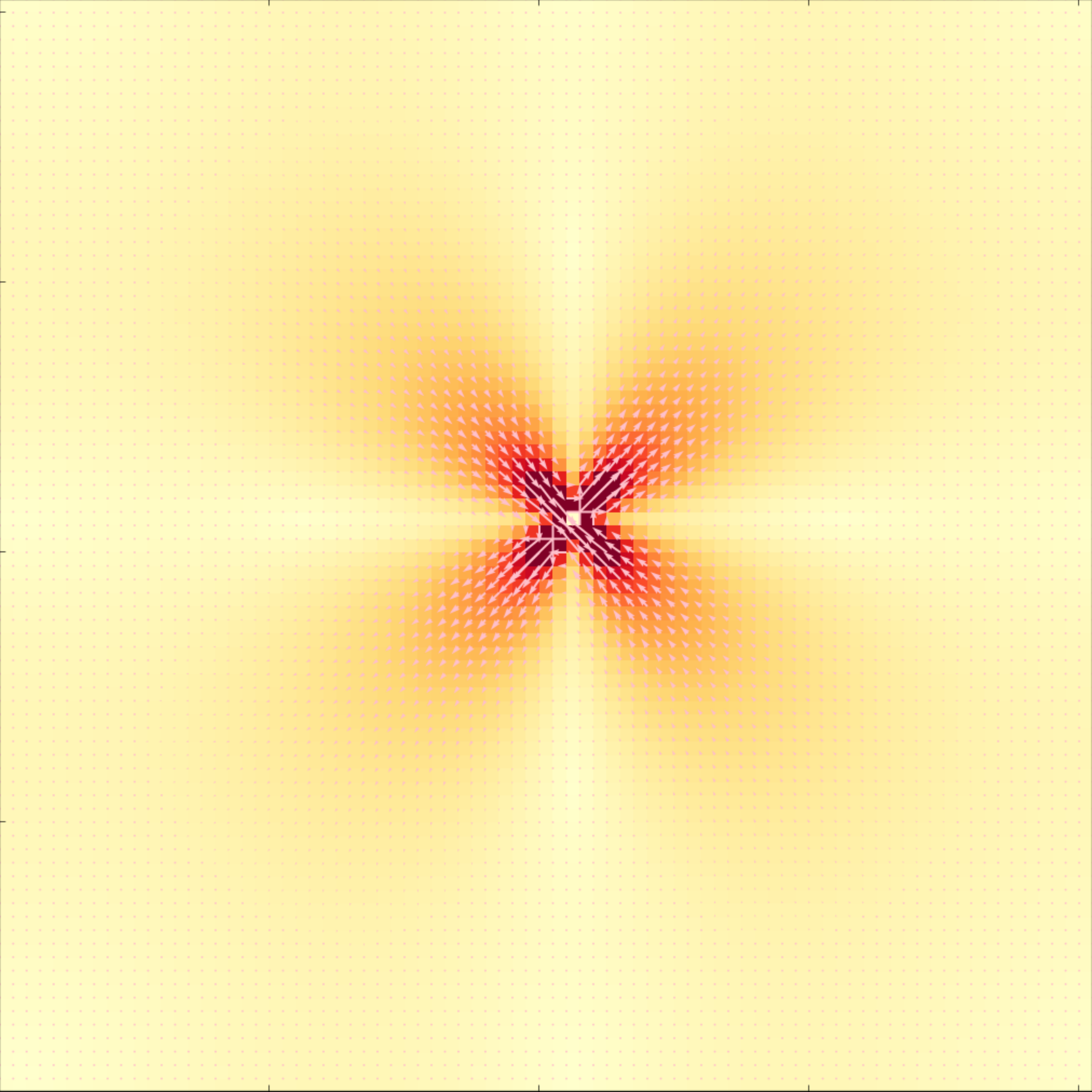}\hspace*{0.3cm}\includegraphics[width=6cm]{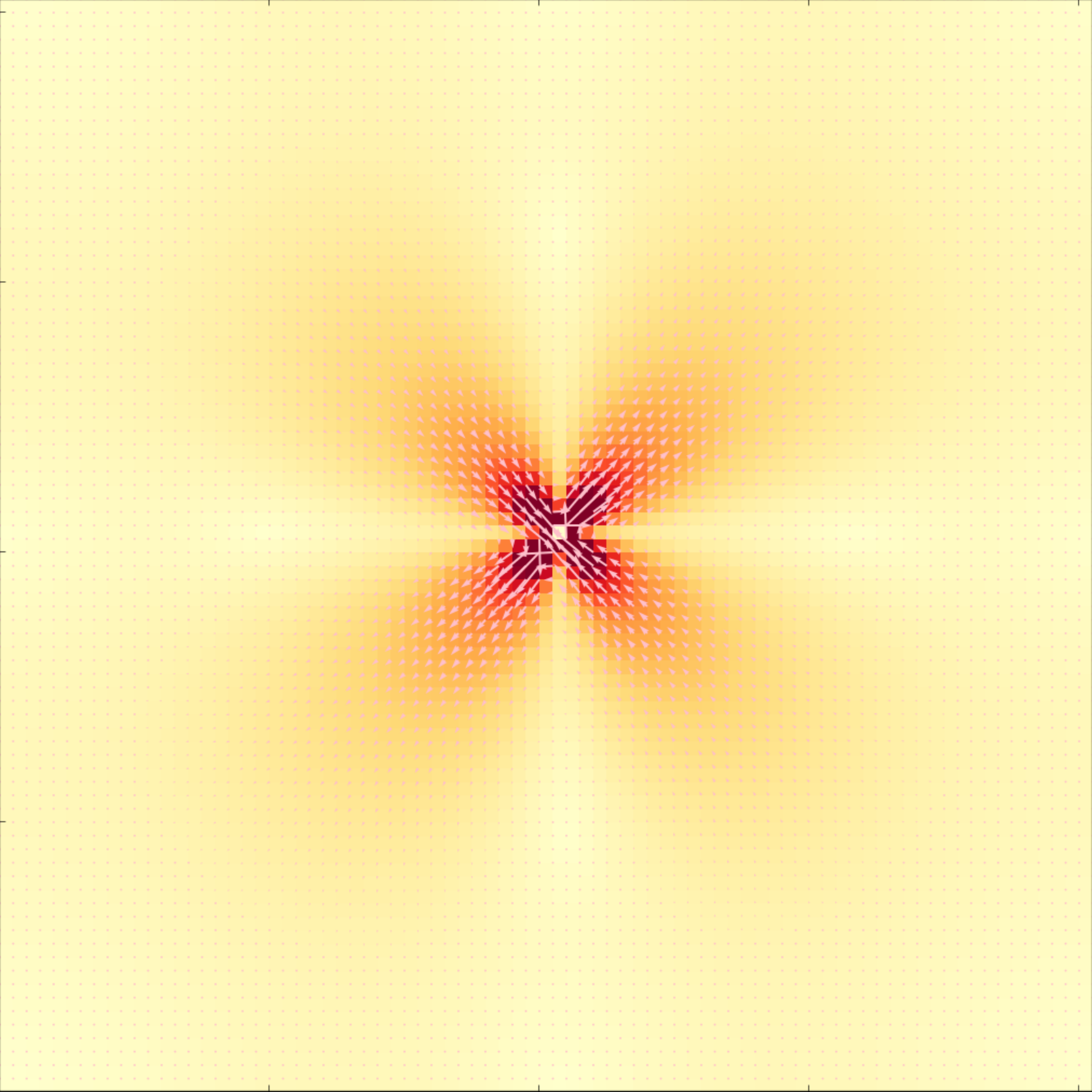}

}
\par\end{centering}

\begin{centering}
\subfloat[$\Delta t=1000$]{\includegraphics[width=6cm]{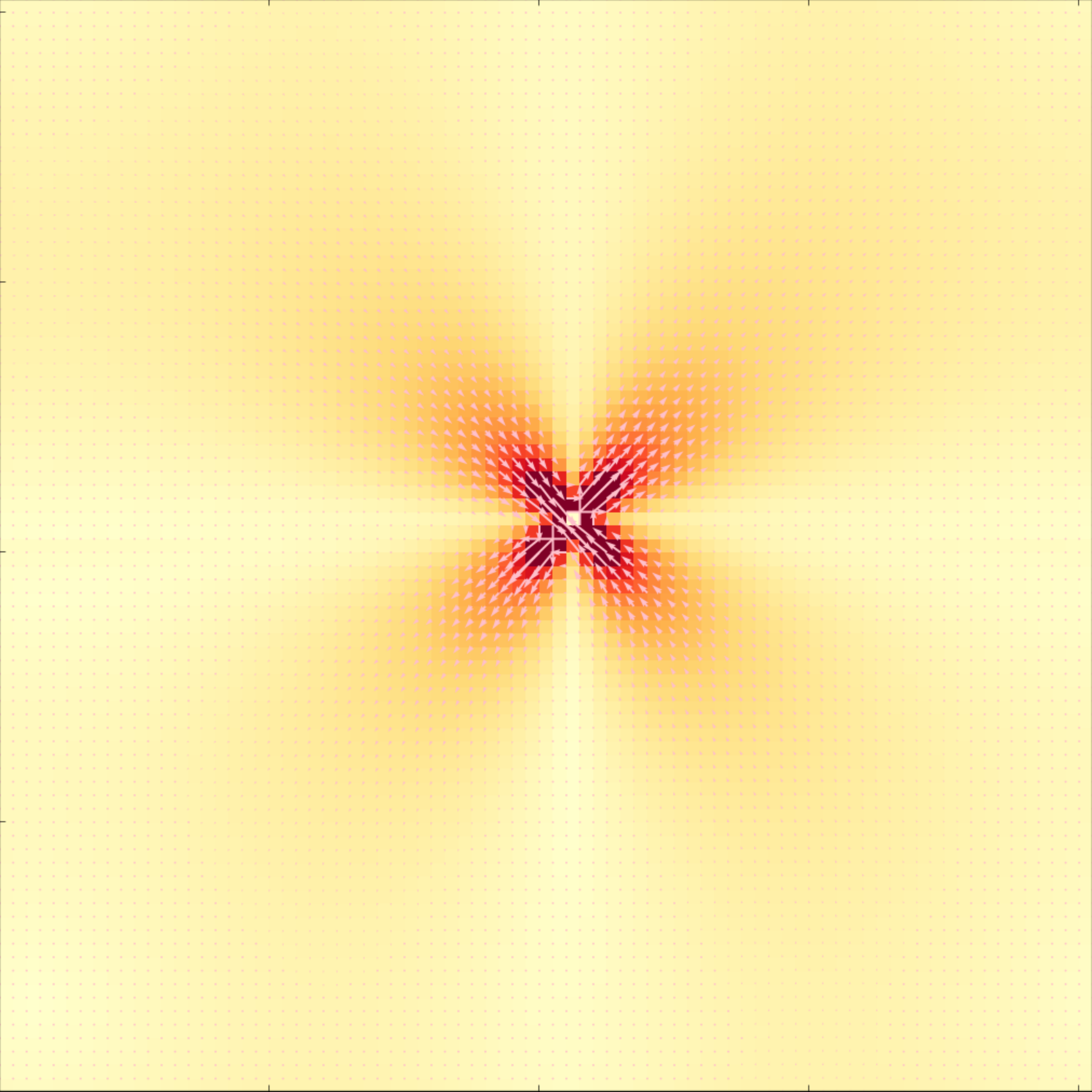}\hspace*{0.3cm}\includegraphics[width=6cm]{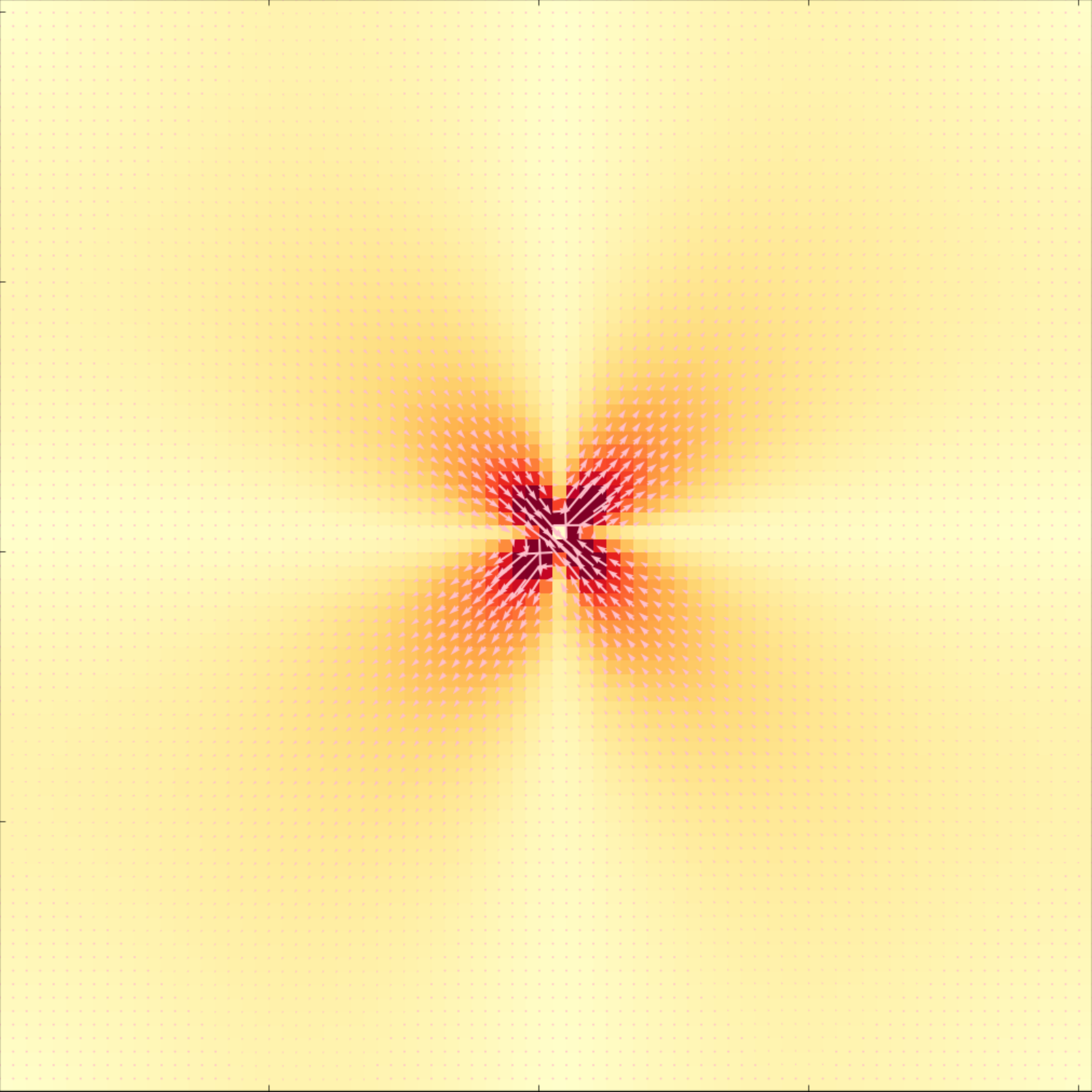}

}
\par\end{centering}

\caption{\label{fig:avg_disp_field_b_FEM}Average displacement field induced
by a shear transformation, after a time lag $\Delta t$, for strong
damping, $\zeta=100$ (hence, $\eta=72.6)$. Refer to Fig.~\ref{fig:avg_disp_field_a_FEM}
for the legend.}
\end{figure}

\selectlanguage{english}%

\subsection{Theoretical rationalisation\label{sub:Theoretical-expectations_prop}}

\selectlanguage{american}%
Puosi and co-workers \citep{Puosi2014} reported that, with a mean-field
dissipative force (\emph{i.e}.,\emph{ }by substituting $\boldsymbol{f_{i}}^{D}=\nicefrac{-m\boldsymbol{v_{i}}}{\tau_{d}}$
for Eq.~\ref{eq:f_diss_DPD} in Eq.~\ref{eq:eq_of_motion_MD}),
\foreignlanguage{english}{$\Delta_{r}(t)$ initially grows in a diffusive
fashion, \emph{i.e.}, }$\Delta_{r}(t)\sim t^{\nicefrac{1}{2}}$, at
large damping, that is to say, for short Langevin damping times $\tau_{d}$.
By contrast, no such diffusive regime is observed here, even for large
damping parameters $\zeta$.\foreignlanguage{english}{ The dissipation
scheme therefore affects the nature of shear wave propagation. Can
this discrepancy be explained theoretically? }

\selectlanguage{english}%

\subsubsection{Mean-field dissipation}

\selectlanguage{american}%
In the presence of a mean-field damping force, force balance on particle
$i$ can schematically be written as
\begin{eqnarray}
m\boldsymbol{\dot{v}_{i}}(t)+\frac{m\boldsymbol{v_{i}}(t)}{\tau_{d}} & \approx & k\sum_{\left\langle j|i\right\rangle }\left(\boldsymbol{u}_{j}(t)-\boldsymbol{u}_{i}(t)\right),\label{eq:MD_prop_FEM}
\end{eqnarray}
where the sum runs over the neighbours $j$ of $i$, $k$ is a typical
stiffness, \emph{i.e.}, the order of magnitude of the relevant Hessian
components $\partial^{2}V/\partial\boldsymbol{r_{i}}\partial\boldsymbol{r_{j}}$,
and the $\boldsymbol{u_{j}}$'s are the displacements with respect
to an equilibrium configuration. Let us now introduce a continuous,
coarse-grained displacement field $\boldsymbol{u}(\boldsymbol{r};t)$
and a typical interparticle distance $a_{0}$, and substitute the
former into Eq.~\ref{eq:MD_prop_FEM}, in the overdamped limit $\tau_{d}\rightarrow0$,
\begin{eqnarray*}
\frac{m}{\tau}\frac{\partial\boldsymbol{u}}{\partial t} & \approx & ka_{0}^{2}\nabla^{2}\boldsymbol{u}.
\end{eqnarray*}
In this regime of negligible inertia, we thus obtain a diffusive equation
for the particle displacements, consistently with the MD observations.

\subsubsection{Dissipative Particle Dynamics}

Very crudely, the DPD equations of motion (Eqs.~\ref{eq:eq_of_motion_MD}-\ref{eq:f_diss_DPD})
are approximated by
\begin{eqnarray}
m\ddot{\boldsymbol{u}} & \approx & \tilde{\zeta}\sum_{\left\langle j|i\right\rangle }\left(\boldsymbol{\dot{u}}_{j}(t)-\boldsymbol{\dot{u}}_{i}(t)\right)+k\sum_{\left\langle j|i\right\rangle }\left(\boldsymbol{u}_{j}(t)-\boldsymbol{u}_{i}(t)\right)\nonumber \\
m\ddot{\boldsymbol{u}} & \approx & \tilde{\zeta}a_{0}^{2}\nabla^{2}\boldsymbol{\dot{u}}+ka_{0}^{2}\nabla^{2}\boldsymbol{u},\label{eq:DPD_eq_FEM}
\end{eqnarray}
where $\tilde{\zeta}\equiv\zeta w^{2}\left(a_{0}\right)$.

Equation~\ref{eq:DPD_eq_FEM} is a diffusion equation (on $\boldsymbol{\dot{u}}$)
\emph{only if} the elastic force is negligible, which will not be
the case in practice. (More generally, Eq.~\ref{eq:DPD_eq_FEM} can
be solved with a space-time Fourier transform, or a joint Laplace-Fourier
transform).

It can also be seen in Eq.~\ref{eq:DPD_eq_FEM} that, regardless
of the value of $\zeta$, the inertial term $m\ddot{\boldsymbol{u}}$
will always dominate at long enough wavelengths. In an unbounded system,
this notably implies that the inertialess Brownian limit, which features
an infinite transverse sound velocity, is singular.

\selectlanguage{english}%

\section{Effect of structural disorder in MD and in FE\label{sec:Disorder_fluctuations_response}}

Let us now investigate the impact of elastic heterogeneity on the
displacement field induced by an individual plastic event, \emph{i.e.},
the importance of fluctuations around the disorder-averaged response. 

\begin{figure}
\begin{centering}
\subfloat[\label{fig:fluct_avg_FEM_1}$\zeta=1$.]{\begin{centering}
\includegraphics[width=7cm]{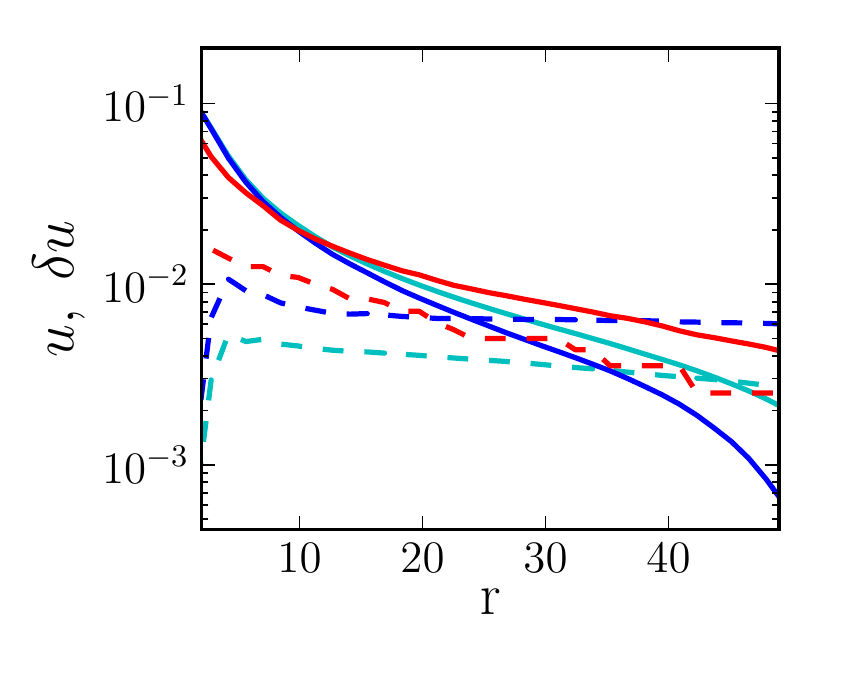}
\par\end{centering}

}\subfloat[\label{fig:fluct_avg_FEM_100}$\zeta=100$.]{\begin{centering}
\includegraphics[width=7cm]{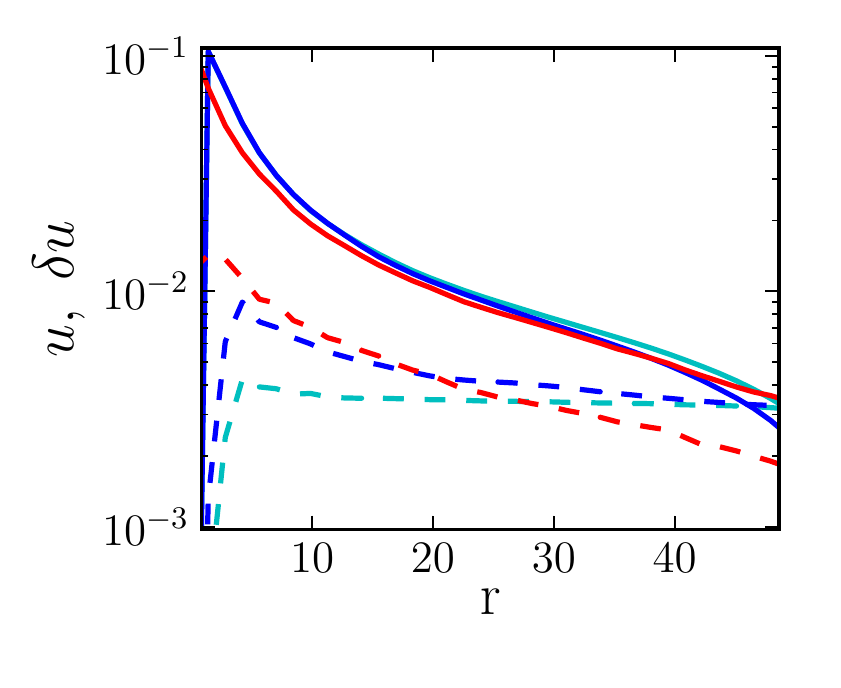}
\par\end{centering}

}
\par\end{centering}

\caption{\label{fig:fluct_avg_FEM}(\emph{Solid line}s) mean value $u$ and
(\emph{dashed lines}) standard deviation $\delta u$ of the displacement
norm along a diagonal axis $\boldsymbol{e}_{\mathrm{diag}}=\frac{\sqrt{2}}{2}\left(\boldsymbol{e_{x}}+\boldsymbol{e_{y}}\right)$,
after a time lag $\Delta t=1000$, as a function of the distance (in
FE units).\protect \\
(\emph{Red}) MD; (\emph{cyan}) FE, het. iso.; (\emph{blue}) FE, het.
aniso.}
\end{figure}

\selectlanguage{american}%
The norm of the average displacement $\boldsymbol{u}\left(\boldsymbol{r};t\right)$
along a diagonal direction, at a long time lag $\Delta t=1000$, is
plotted in Fig.~\ref{fig:fluct_avg_FEM} for $\zeta=1$ and $\zeta=100$,
along with the associated standard deviation $\delta u$, \emph{i.e.,}
\[
\delta u\left(\boldsymbol{r};t\right)=\sqrt{\left\langle \left[\boldsymbol{u}^{(d)}\left(\boldsymbol{r};t\right)-\boldsymbol{u}\left(\boldsymbol{r};t\right)\right]^{2}\right\rangle _{d}},
\]
where the brackets denote an average over the realisations of disorder.
Incidentally, one may notice that, for $\zeta=1$ (Fig.~\ref{fig:fluct_avg_FEM_1}),
MD and FE do not coincide satisfactorily with respect to the average
displacements, but this is mostly due to a loss of synchronization:
the oscillations described in Section~\ref{sec:propagation_comp}
have not died out yet at this time lag and they are not exactly in
phase in the different systems. Had the true steady-state limit, $\Delta t\rightarrow\infty$,
been reached (at the expense of much longer simulations), we would
have expected much better agreement on the average displacements.
This expectation is supported by the coincidence of the average displacements
at $\Delta t=1000$ under strong damping, for $\zeta=100$ (see Fig.~\ref{fig:fluct_avg_FEM_100}),
in which case dissipation is more efficient and the steady state is
reached after fewer MD steps; indeed, in the linear regime probed
here, the final state should be independent of the dynamics, hence
of $\zeta$.

Regarding the fluctuations, the main result is that their order of
magnitude is well reproduced by the FE simulations, both with isotropic
blocks (het. iso., $\mu_{1}=\mu_{2}$) and with anisotropic blocks
(het. aniso.), although, quite naturally, het. aniso. displays larger
fluctuations than het. iso. Moreover, it is noteworthy that these
corrections $\delta u$ are roughly half as large as the mean reponse
at a distance of, \emph{e.g.}, $50\sigma_{AA}$. To avoid any misunderstanding
on the possible nature of the fluctuations measured in MD, let us
recall here that the centre of mass of the MD simulation cell is kept
fixed, which prevents the variable global translations of the system
that are sometimes observed otherwise (and which then dominate the
fluctuations)%
\footnote{When the centre of mass of the MD simulation cell is not kept fixed,
the fluctuations $\delta u$ measured in MD are significantly larger
and their profile with respect to the distance $r$ to the origin
(dashed lines in Fig.~\ref{fig:fluct_avg_FEM}) is almost flat.%
}. 

\begin{figure}
\begin{centering}
\subfloat[$\Delta t=10$]{\begin{centering}
\includegraphics[width=7cm]{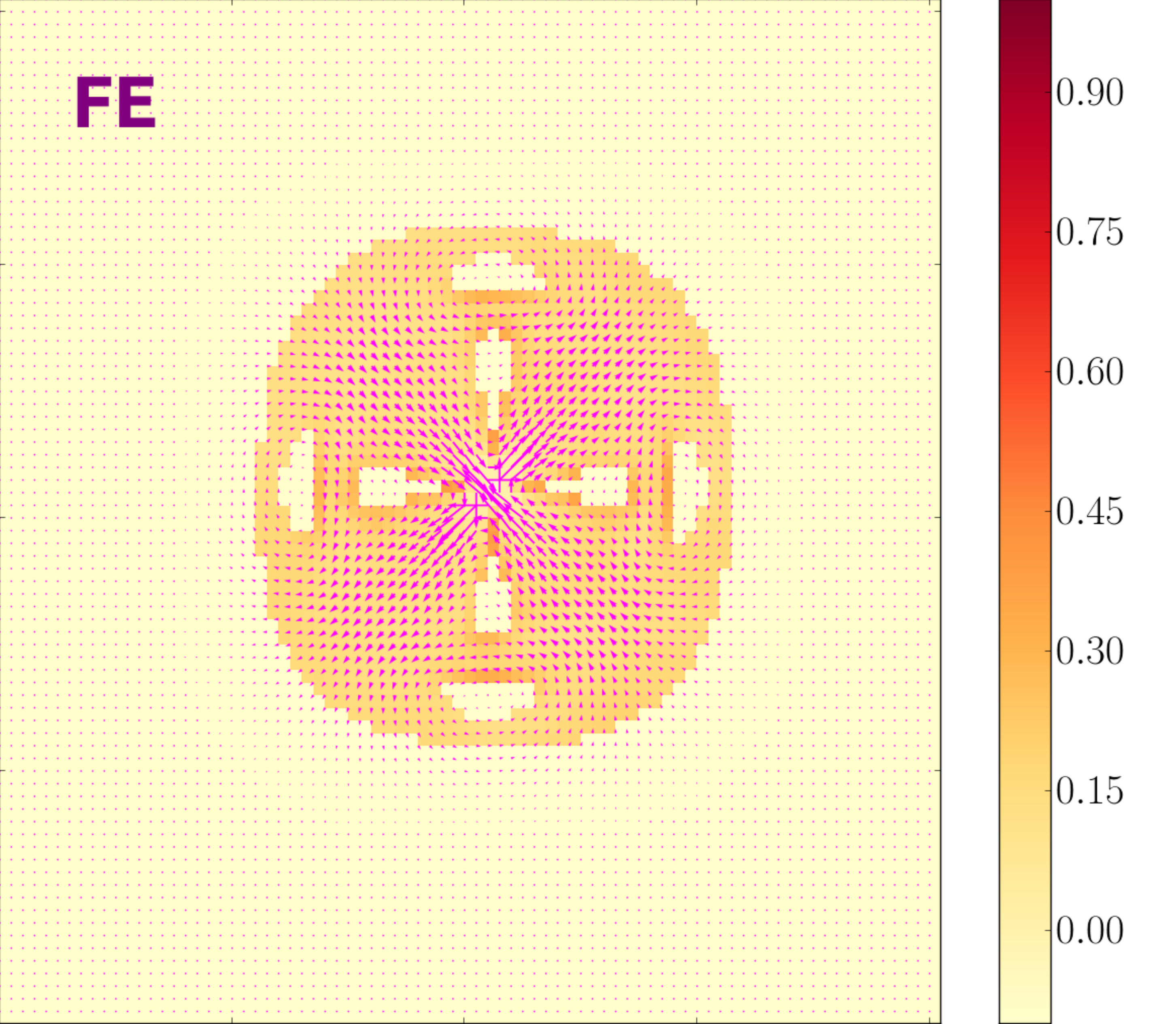}\includegraphics[width=7cm]{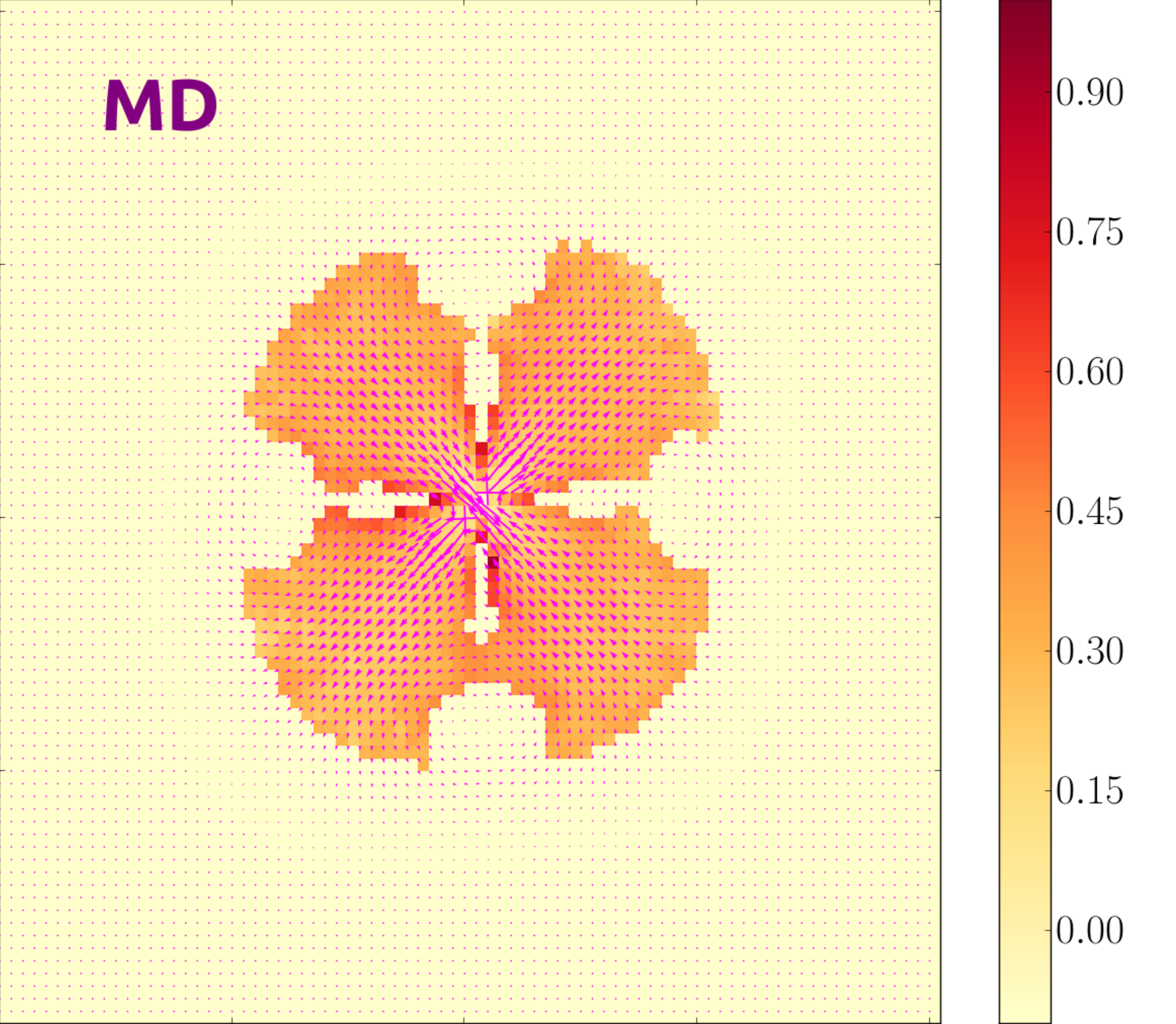}
\par\end{centering}

}
\par\end{centering}

\begin{centering}
\subfloat[$\Delta t=100$]{\begin{centering}
\includegraphics[width=7cm]{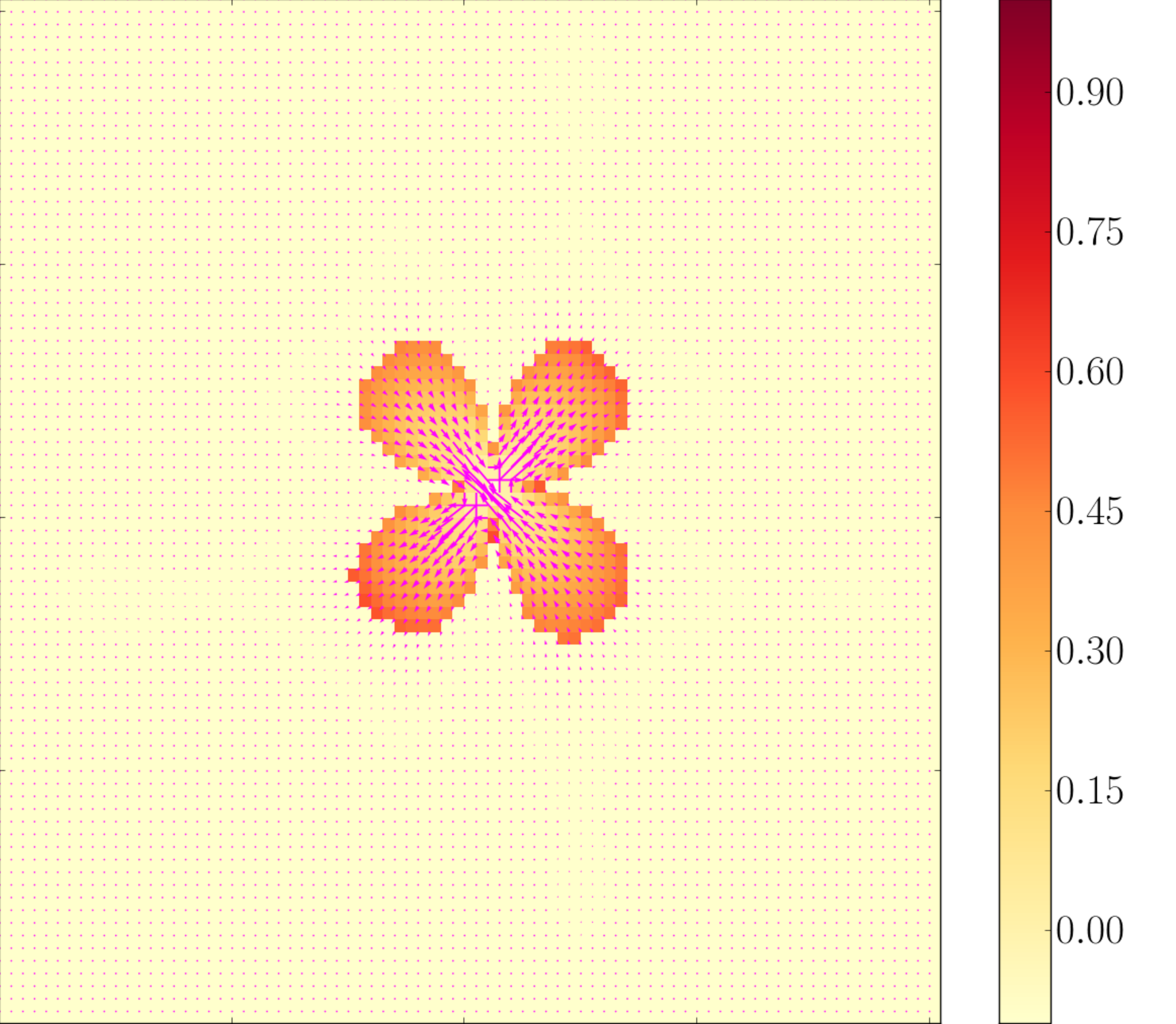}\includegraphics[width=7cm]{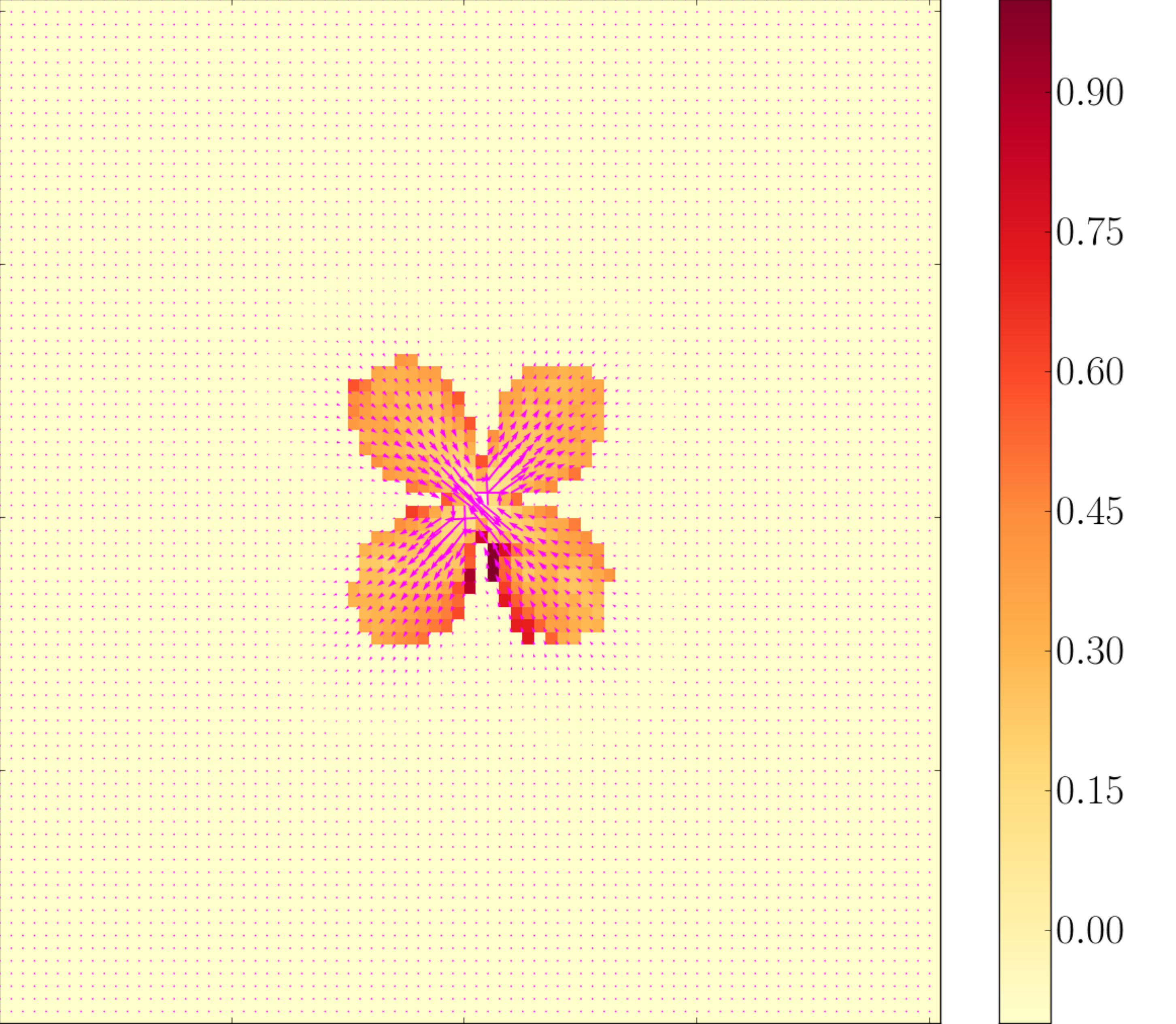}
\par\end{centering}

}
\par\end{centering}

\begin{centering}
\subfloat[$\Delta t=1000$]{\begin{centering}
\includegraphics[width=7cm]{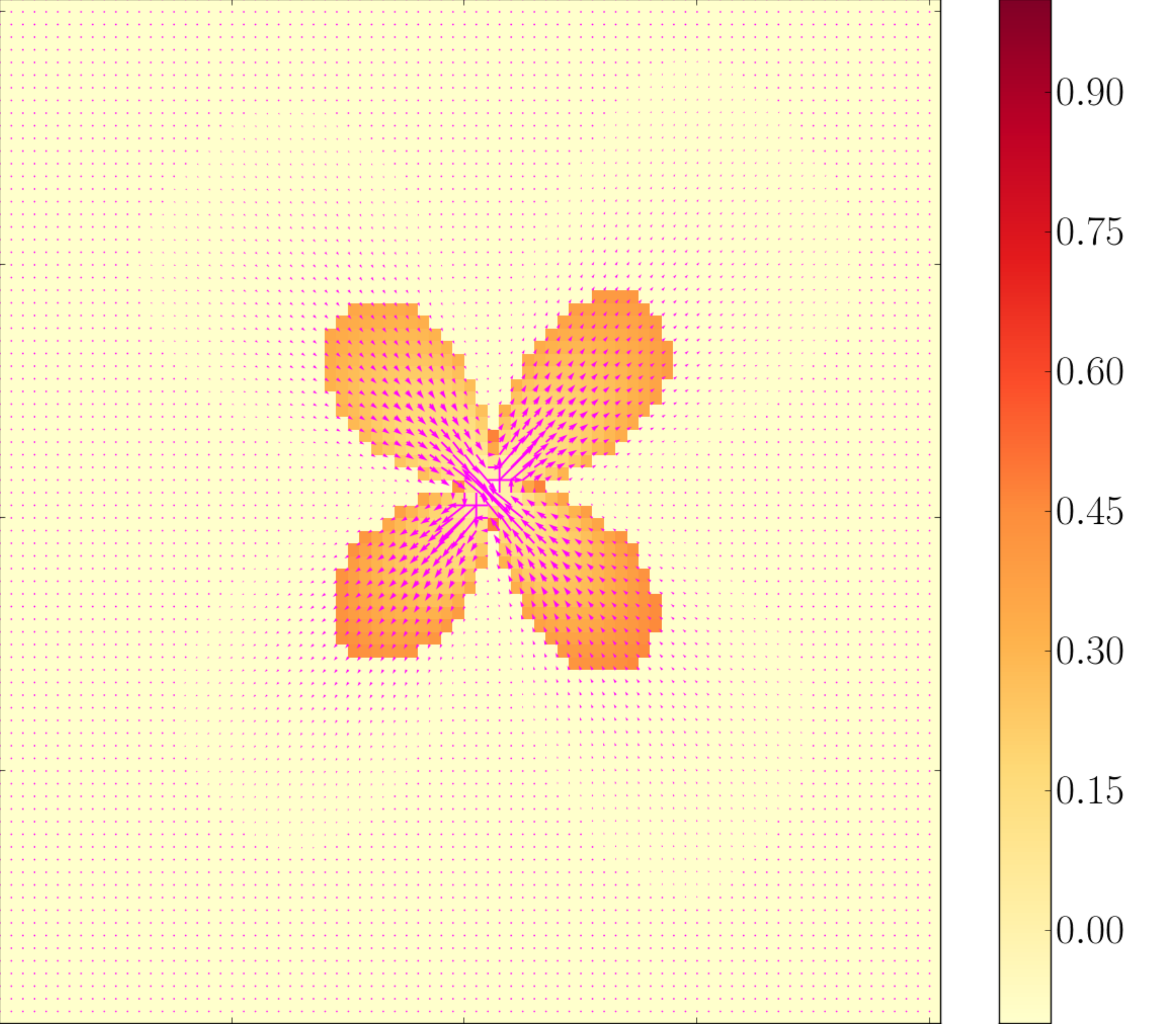}\includegraphics[width=7cm]{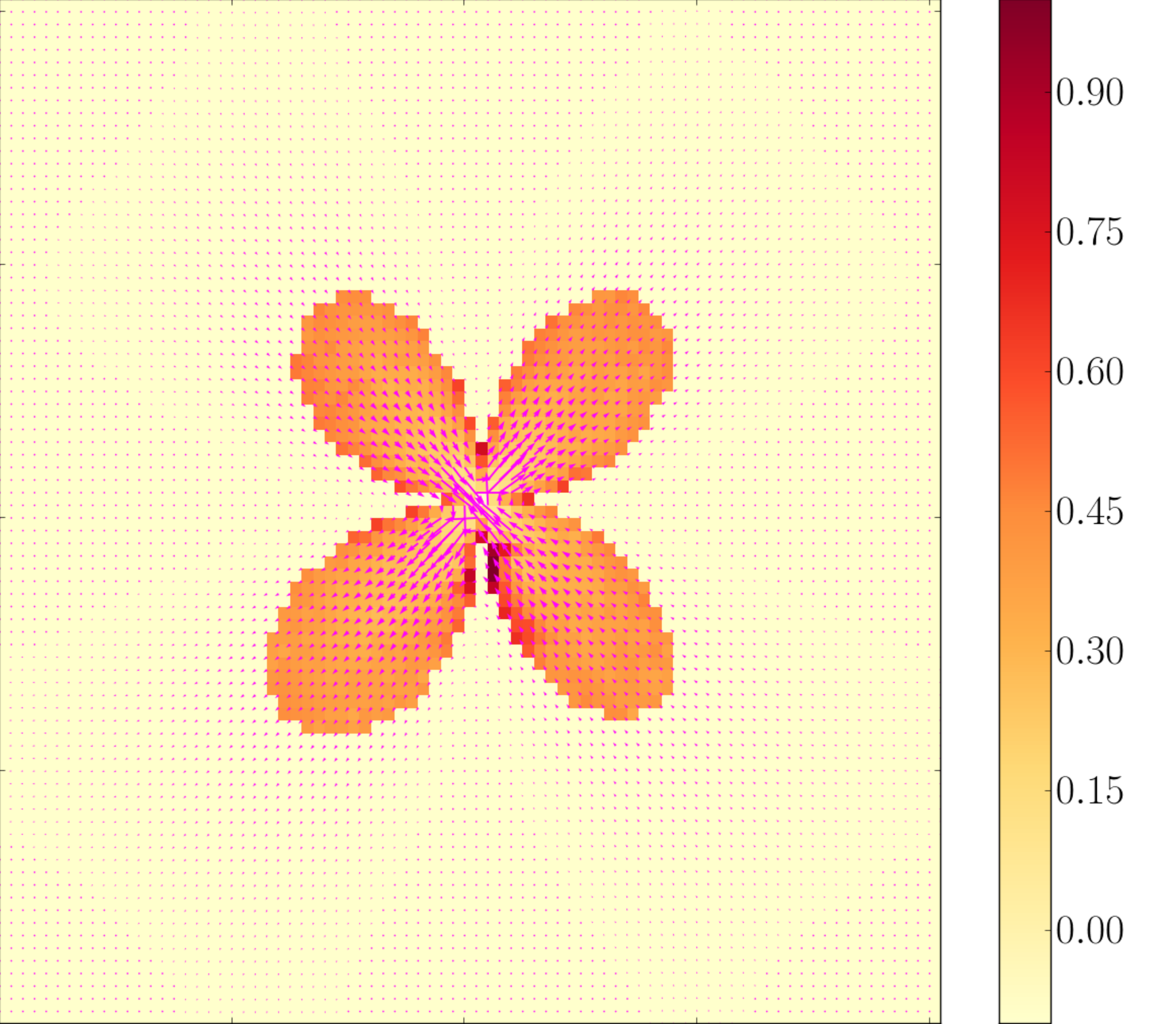}
\par\end{centering}

}
\par\end{centering}

\caption{\label{fig:fluct_over_mean_FEM}Colour map of the relative displacement
norm fluctuations $\delta u(\boldsymbol{r};t)/u(\boldsymbol{r};t)$
for $\zeta=1$. The regions where $u(\boldsymbol{r};t)<10^{-2}$ are
overlaid in light yellow.}
\end{figure}

With regard to the spatial distribution of $\delta u$, colour maps
of the relative fluctuations $\delta u(\boldsymbol{r};t)/u(\boldsymbol{r};t)$
are presented in Fig.~\ref{fig:fluct_over_mean_FEM}. In regions
with non-negligible displacements, \emph{i.e.}, $u(\boldsymbol{r};t)\geqslant10^{-2}$,
the relative fluctuations are approximately homogeneous and tend to
increase slightly with time.

\selectlanguage{english}%
In conclusion to this section, taking into account the broad distribution
of shear moduli in FE has enabled us to recover the fluctuations observed
in MD. This further confirms the role of structural disorder on the
redistribution of stress induced by a plastic event. In the last section,
we go one step further by attempting to reproduce the individual,
time-dependent response to a \emph{given} plastic event in MD with
the simple FE framework.

\section{Time-dependent response to a particular plastic event}

Even though the study of the propagation dynamics (Section~\ref{sec:Disorder-averaged-propagation})
and of disorder-induced fluctuations (Section~\ref{sec:Disorder_fluctuations_response})
validates the FE method for (future) use in, \emph{e.g.}, mesoscopic
rheological models, we would like to know whether the comparison can
be pushed further. More precisely, can the FE routine describe the
details of the elastic response in a \emph{particular }configuration?

\begin{figure}
\selectlanguage{american}%
\begin{centering}
\subfloat[$\Delta t=1$]{\includegraphics[width=7cm]{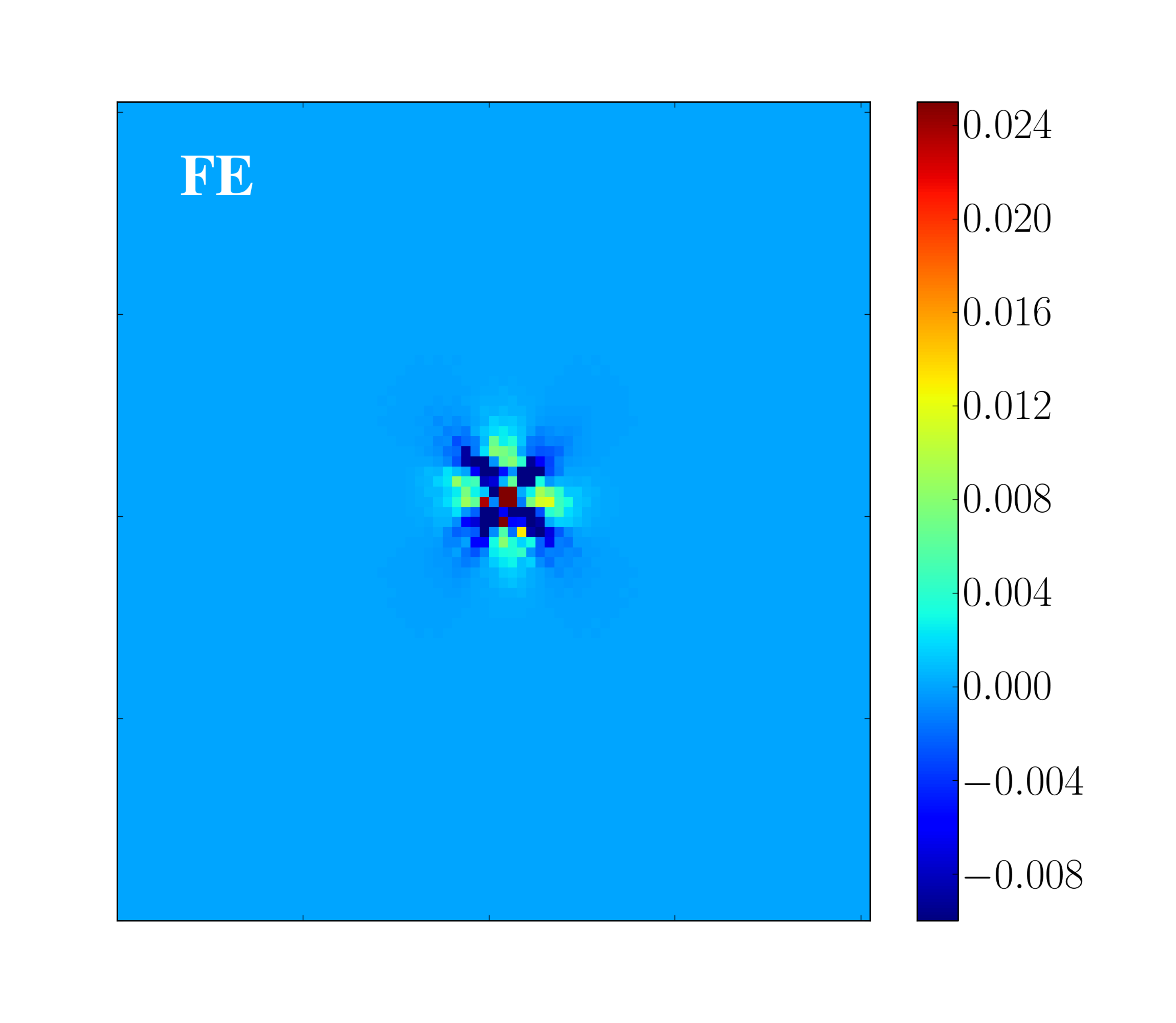}

\includegraphics[width=7cm]{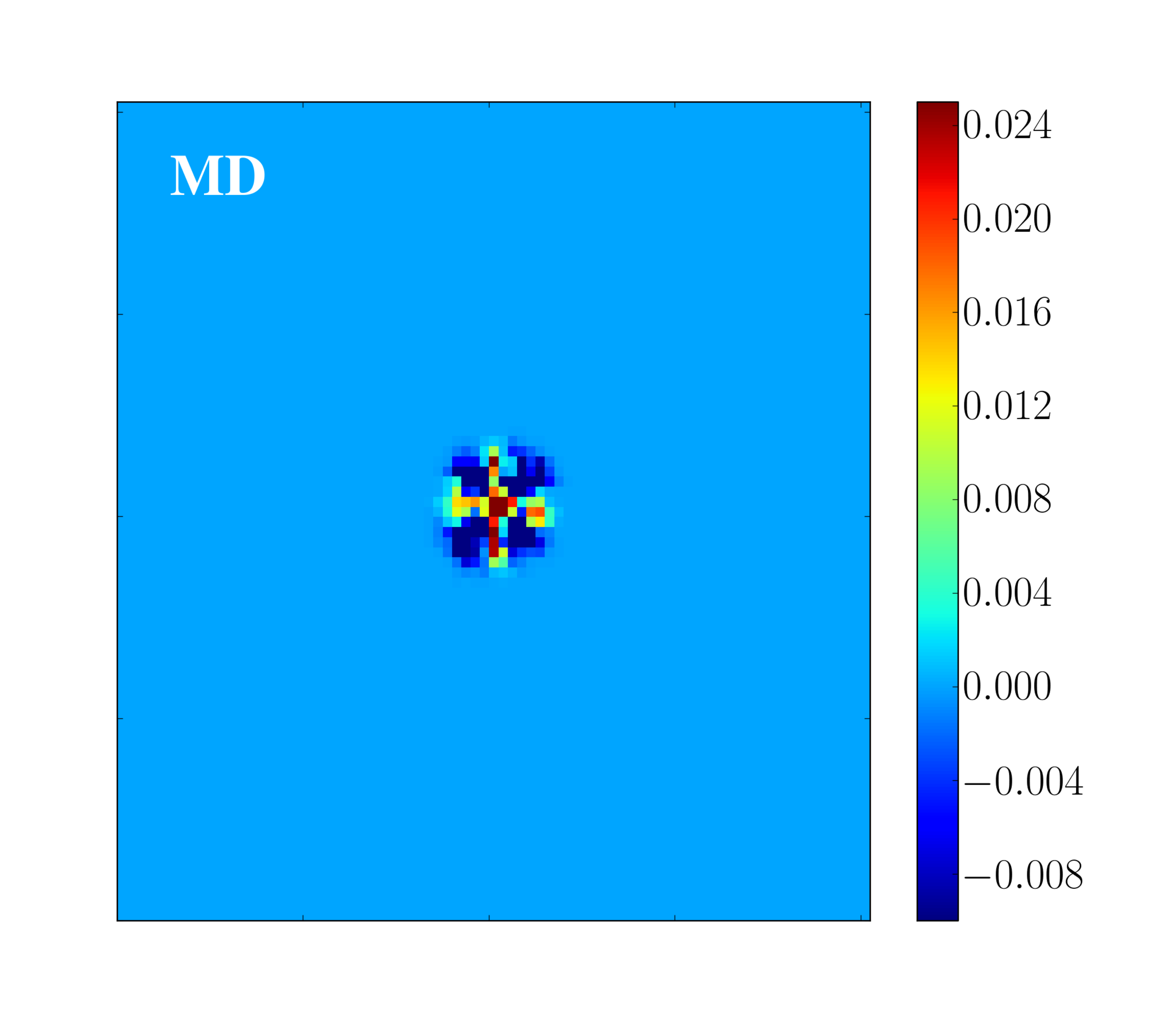}

}
\par\end{centering}

\begin{centering}
\subfloat[$\Delta t=10$]{\includegraphics[width=7cm]{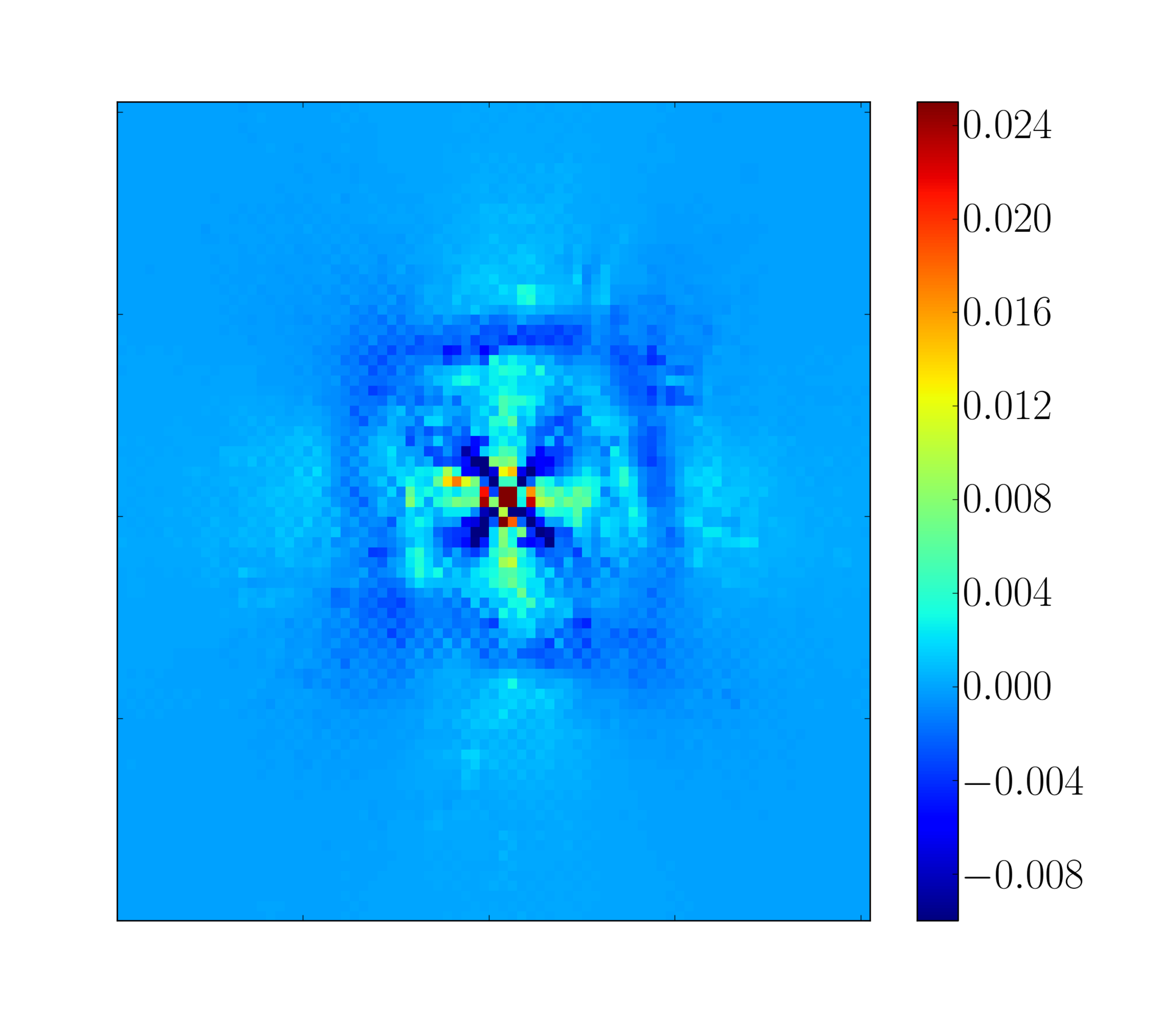}

\includegraphics[width=7cm]{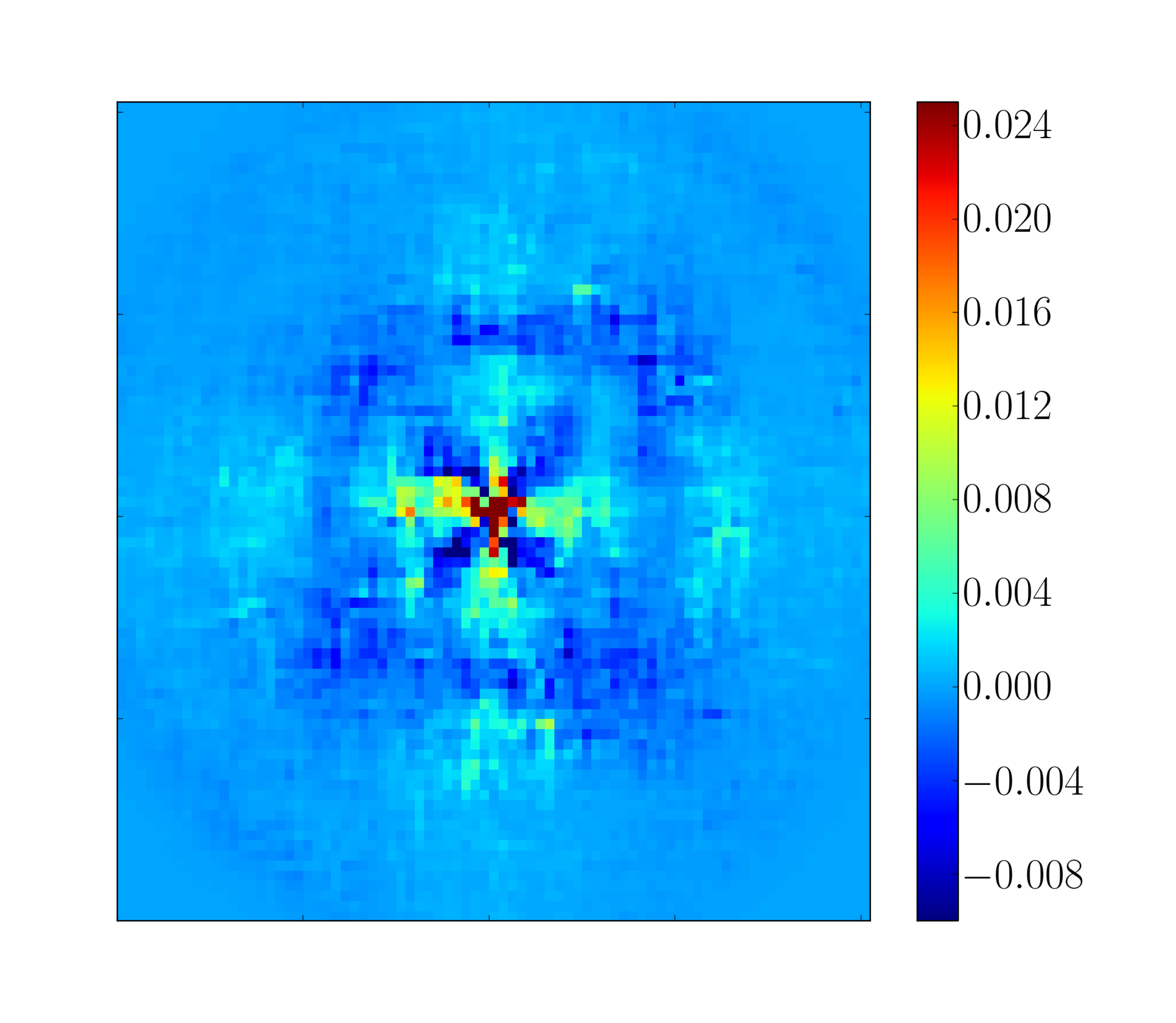}

}
\par\end{centering}

\begin{centering}
\subfloat[$\Delta t=100$]{\includegraphics[width=7cm]{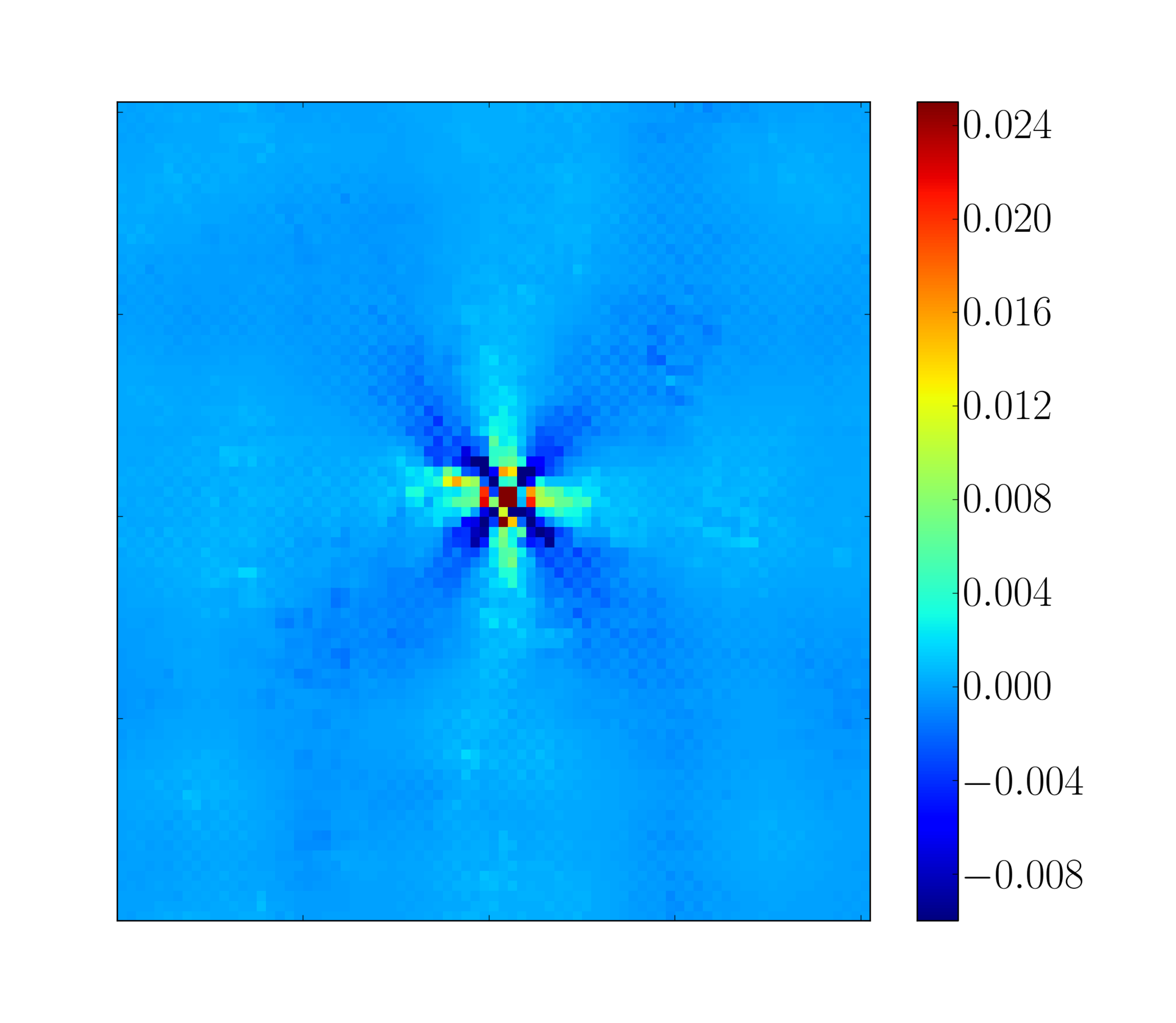}

\includegraphics[width=7cm]{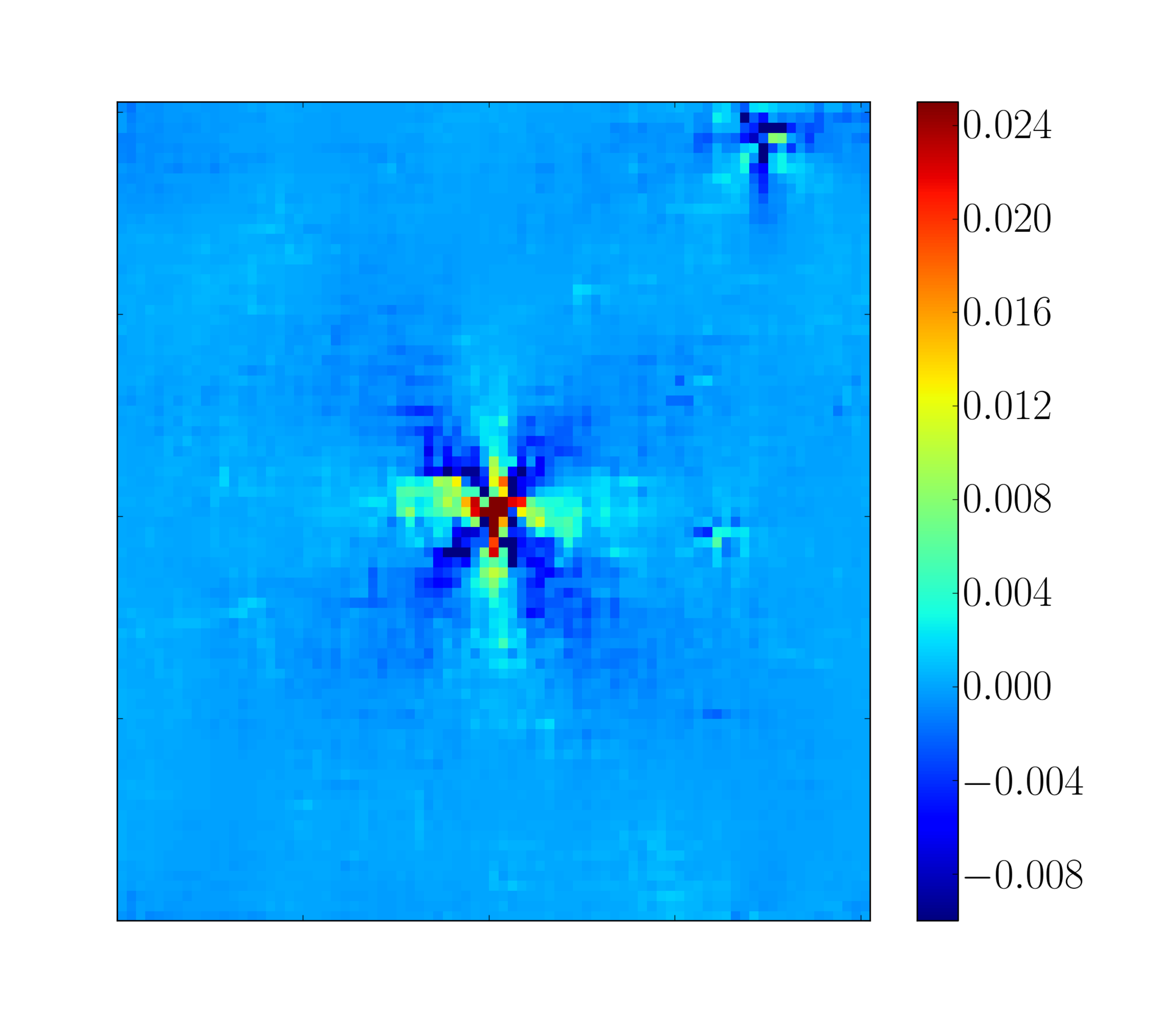}

}
\par\end{centering}

\selectlanguage{english}%
\caption{\selectlanguage{american}%
\label{fig:singleST_FEM}Local strain field induced by a particular
shear transformation at different lag times, for $\zeta=1$.\protect \\
(\emph{Left}) FE, with an elastic configuration modelled on the MD
system; (\emph{right}) MD.\selectlanguage{english}
}
\end{figure}

To address this question, within the third type of FE mode, namely,
het. aniso., the local shear moduli $\mu_{1}$ and $\mu_{2}$ and
the angle $\theta$ of each FE macro-element (\emph{i.e., }set of
four adjacent elements) are directly extracted from the corresponding
region in the MD system. Then, we compute the coarse-grained strain
field%
\footnote{In MD, local strains are computed after coarse-graining the displacement
field on a grid similar to the FE one; note that the strain field
is expected to be less sensitive to heterogeneities than the displacement
field.%
} induced by shear transformations occurring at given position in the
sample, an example of which is shown in Fig.~\ref{fig:singleST_FEM}.

Clearly, the MD response and its FE counterpart look alike and both
exhibit the distinctive quadrupolar angular structure associated to
the response in a uniform medium. However, are the disorder-induced
fluctuations, \emph{i.e}, the deviations from this average response,
also similar in MD and FE? In an endeavour to answer this question,
we have looked at the \emph{deviations} in half a dozen particular
configurations (\emph{not shown}) and considered a couple of basic
measures of similarity, but our results remain inconclusive in this
respect: there is no quantifiable evidence that the disorder-induced
fluctuations in a particular MD configuration are satisfactorily reproduced
in FE.

\section{Conclusions}

In conclusion, we have extracted information about the local elastic
constants of a binary Lennard-Jones mixture and the viscosity associated
with a DPD damping scheme. Consistently with the findings of \citet{Mizuno2013moduli},
we have found that \foreignlanguage{american}{the local shear moduli
are more broadly distributed (on a relative basis) than local bulk
moduli.}

These elastic and viscous properties were used as input in a simple
FE routine and an ideal shear transformation was artificially triggered
in the (FE and MD) systems.

\selectlanguage{american}%
We observed that the \emph{average} time-dependent elastic response
to this transformation in a disordered medium is similar to the propagation
in a uniform medium and it is well reproduced in the FE simulations.
However, fluctuations with respect to the average displacement field
are considerable, with relative fluctuations of a few tens of percents.
The approximate magnitude of these fluctuations is captured by FE
simulations on heterogeneous, but locally isotropic systems. Refining
the description by considering the elastic anisotropy on the mesoscale
does not play a major role in this respect.

It should however be stressed that, throughout our investigation,
shear transformations were arbitrarily imposed, through an instantaneous
displacement of particles (or FE nodes). However, in a \emph{bona
fide} simulation, the dynamics of shear transformations are determined
by the system itself; two dynamical regimes can then be envisioned:

(i) if inertia is negligible, the competitition between elasticity
and viscosity sets the timescale of the rearrangement, $\tau=\eta/\mu$,

(ii) if the rearrangement mostly consists in the damping of the inertial
force (initially generated by elasticity), then the duration of a
rearrangement is set by the inverse damping coefficient $\zeta^{-1}$.

\selectlanguage{english}%
All in all, our method represents a powerful new framework for rheological
models for amorphous solids, which improves on the traditional use
of an analytical elastic propagator and the computation of the response
by means of a Fast Fourier Transform, in that it accounts for structural
disorder and inertial effects, whose impact has been underscored by
\citet{Salerno2012}, it can be extended to arbitrary (in particular,
confined) geometries, and it may include pre-existing local defects
in the material, such as cracks. A further asset of this strategy
is that, notwithstanding the enhanced capabilities of the algorithm,
its complexity in terms of number of operations scales linearly with
the number of blocks (or FE cells) for large systems, that is, with
a scaling comparable to that of the Fast Fourier Transform routine.

\bigskip{}

\emph{Acknowledgements}

AN thanks Richard \noun{Michel} for his help with the Finite Element
method. The MD simulations were carried out on clusters belonging
to the CIMENT infrastructure (https://ciment.ujf-grenoble.fr), which
is supported by the Rhône-Alpes region (GRANT CPER07\_13 CIRA: http://www.ci-ra.org),
using LAMMPS molecular dynamics software \citep{Plimpton1995} (http:
//lammps.sandia.gov). JLB is supported by Institut Universitaire de
France and by grant ERC-2011-ADG20110209.

\bigskip{}

\emph{Bibliography}

\appendix

\section{Simplified Finite Element routine\label{app:FE_routine}}

Bearing in mind our pursuit of minimalism, we choose a simple regular
square meshgrid, as sketched in Fig.~\ref{fig:mesh_FEM}. If one
assumes that the strain and stress fields are approximately uniform
in each element, the following equations can be written between the
(nodal) displacements $\left(u_{x},u_{y}\right)$ and the (elemental)
strains $\boldsymbol{\epsilon}$, on the one hand, and the (nodal)
forces $\left(f_{x}^{\mathrm{el}},f_{y}^{\mathrm{el}}\right)$ and
the (elemental) stresses $\boldsymbol{\sigma}$, on the other hand:

\begin{equation}
\boldsymbol{\epsilon}={\bf B}\cdot\left(\begin{array}{c}
u_{x}^{(0)}\\
u_{y}^{(0)}\\
\vdots\\
u_{x}^{(3)}\\
u_{y}^{(3)}
\end{array}\right)\text{ and }\boldsymbol{\sigma}=-{\bf B}\cdot\left(\begin{array}{c}
f_{x}^{\mathrm{el}\,(0)}\\
f_{y}^{\mathrm{el}\,(0)}\\
\vdots\\
f_{x}^{\mathrm{el}\,(3)}\\
f_{y}^{\mathrm{el}\,(3)}
\end{array}\right),\label{eq:B_transp_matrix_FEM}
\end{equation}
where the nodes of the element have been numbered from 0 to 3 counter-clockwise,
starting from the bottom left corner, \emph{viz.}, \foreignlanguage{french}{${3\atop 0}{\scriptstyle \square}{2\atop 1}$},
and $u_{x}^{(0)}$ denotes the displacement along $x$ at the (0)
node\emph{, etc.} Here, we have used condensed notations for the 2D
strains and the stresses, \emph{viz.}, 
\[
\boldsymbol{\epsilon}\equiv\left(\begin{array}{c}
\epsilon_{xx}\\
\epsilon_{yy}\\
\sqrt{2}\epsilon_{xy}
\end{array}\right)\text{ and }\boldsymbol{\sigma}\equiv\left(\begin{array}{c}
\sigma_{xx}^{\mathrm{el}}\\
\sigma_{yy}^{\mathrm{el}}\\
\sqrt{2}\sigma_{xy}^{\mathrm{el}}
\end{array}\right),
\]
and the matrix ${\bf B}$ is given by \foreignlanguage{french}{
\[
{\bf B}\equiv\nicefrac{1}{2}\left[\begin{array}{cccccccc}
-1 &  & 1 &  & 1 &  & -1\\
 & -1 &  & -1 &  & 1 &  & 1\\
\nicefrac{-1}{\sqrt{2}} & \nicefrac{-1}{\sqrt{2}} & \nicefrac{-1}{\sqrt{2}} & \nicefrac{1}{\sqrt{2}} & \nicefrac{1}{\sqrt{2}} & \nicefrac{1}{\sqrt{2}} & \nicefrac{1}{\sqrt{2}} & \nicefrac{-1}{\sqrt{2}}
\end{array}\right].
\]
}

\selectlanguage{french}%
Notice that our simplified FE method is close to a Finite Volume method,
in practice. \foreignlanguage{english}{The $\sqrt{2}$ prefactors
have been introduced with foresight (see Section~\ref{sec:local_el_csts})
and the {}``minus'' sign preceding ${\bf B}$ in Eq.~\ref{eq:B_transp_matrix_FEM}
should not come as a surprise if one recalls that $\boldsymbol{f}^{\mathrm{el}\,(i)}$
is the force exerted \emph{by} the element \emph{on} node $i$.}

\selectlanguage{english}%
Contrary to traditional FE codes, the mesh will here remain static,
\emph{i.e.}, not be distorted owing to the material deformation.

\selectlanguage{french}%

\subsection{Elastic force-displacement matrix\label{sec:elastic_force_disp_matrix_FEM}}

The objective is now to rewrite Eq.~\ref{eq:Continuum} in terms
of nodal displacements and forces in order to arrive at Eq.~\ref{eq:Discrete2}.

\selectlanguage{english}%
To relate the nodal displacements and the nodal forces in each element,
we make use of the constitutive equation of the material. 

To start with, the elastic contribution is governed by Hooke's law,
which reads, in condensed notations \citep{Tsamados2009},
\begin{equation}
\boldsymbol{\sigma}=\mathbf{C}\cdot\boldsymbol{\epsilon},\label{eq:Stiffness_0}
\end{equation}
where ${\bf C}$ is a $3\times3$ real matrix. Substituting from Eq.~\ref{eq:B_transp_matrix_FEM},
one obtains the local relation between the forces exerted on the nodes
by the material element under consideration and the displacements
at the nodes, \emph{viz.}, 
\begin{equation}
\left(\begin{array}{c}
f_{x}^{\mathrm{el}\,(0)}\\
f_{y}^{\mathrm{el}\,(0)}\\
\vdots\\
f_{x}^{\mathrm{el}\,(3)}\\
f_{y}^{\mathrm{el}\,(3)}
\end{array}\right)=-{\bf B}^{\top}\mathbf{C}{\bf B}\cdot\left(\begin{array}{c}
u_{x}^{(0)}\\
u_{y}^{(0)}\\
\vdots\\
u_{x}^{(3)}\\
u_{y}^{(3)}
\end{array}\right).\label{eq:BCB_FEM}
\end{equation}

\selectlanguage{american}%
\smallskip{}

\selectlanguage{english}%
To proceed, the \emph{local} elastic force-displacement matrices ${\bf K}\equiv-{\bf B}^{\top}\mathbf{C}{\bf B}$
are assembled into a global elastic force-displacement matrix $\boldsymbol{\mathcal{K}}$,~\emph{viz.,}
\[
\left(\begin{array}{c}
f_{x}^{\mathrm{el}\,\mathbf{(N-1)}}\\
f_{y}^{\mathrm{el}\,\mathbf{(N-1)}}\\
\vdots\\
f_{x}^{\mathrm{el}\,\mathbf{(0)}}\\
f_{y}^{\mathrm{el}\,{\bf (0)}}
\end{array}\right)=\boldsymbol{\mathcal{K}}\cdot\left(\begin{array}{c}
u_{x}^{{\bf (N-1)}}\\
u_{y}^{{\bf (N-1)}}\\
\vdots\\
u_{x}^{{\bf (0)}}\\
u_{y}^{{\bf (0)}}
\end{array}\right),
\]
where the bold superscripts refer to the global labels used in Fig.~\ref{fig:mesh_FEM},
by opposition with the elemental labels used in Eq.~\ref{eq:BCB_FEM}.
Here, $\boldsymbol{\mathcal{K}}$ is a sparse $2N\times2N$ matrix.

\subsection{Viscous force-velocity matrix}

The foregoing derivation relies on the linear relation connecting
local strains and elastic stresses. Thus, it can straightforwardly
be extended to the viscous stresses, insofar as they are linearly
related with the local strain rates, \emph{viz.}, 
\begin{equation}
\boldsymbol{\dot{\sigma}}^{\mathrm{diss}}=\mathbf{C^{diss}}\cdot\boldsymbol{\dot{\epsilon}}.\label{eq:C_diss_0}
\end{equation}
Globally, the viscous force-velocity relation reads

\[
\left(\begin{array}{c}
f_{x}^{\mathrm{diss}\,\mathbf{(N-1)}}\\
f_{y}^{\mathrm{diss}\,\mathbf{(N-1)}}\\
\vdots\\
f_{x}^{\mathrm{diss}\,\mathbf{(0)}}\\
f_{y}^{\mathrm{diss}\,{\bf (0)}}
\end{array}\right)=\boldsymbol{\mathcal{H}}\cdot\left(\begin{array}{c}
\dot{u}_{x}^{{\bf (N-1)}}\\
\dot{u}_{y}^{{\bf (N-1)}}\\
\vdots\\
\dot{u}_{x}^{{\bf (0)}}\\
\dot{u}_{y}^{{\bf (0)}}
\end{array}\right),
\]
where the $2N\times2N$ matrix $\boldsymbol{\mathcal{H}}$ has been
assembled from elemental matrices of the form $-{\bf B}^{\top}\mathbf{C^{diss}}{\bf B}$.

\subsection{Inertial force-acceleration matrix}

Finally, we must express the inertial forces, that is to say, the
matrix $\boldsymbol{\mathcal{M}}$ in Eq.~\ref{eq:Discrete2}. The
convected part of the material derivative of the velocity, namely,
$\boldsymbol{v}\cdot\left(\nabla\boldsymbol{v}\right)$ , which scales
with $v^{2}$ for elements of unit size, is neglected.

We compute the inertial forces directly at the nodes. In other words,
each node is assigned a mass $m_{0}\equiv\rho V_{0}$, where $V_{0}$
is the elemental volume (\emph{i.e.}, area). Accordingly, the lumped-mass
matrix $\boldsymbol{\mathcal{M}}$ connecting the accelerations at
the nodes to the inertial forces at the nodes is a $2N\times2N$ matrix
with $m_{0}$ on the diagonal, \emph{i.e.}, 
\[
\boldsymbol{\mathcal{M}}=\left(\begin{array}{ccc}
m_{0}\\
 & \ddots\\
 &  & m_{0}
\end{array}\right).
\]

Below, we detail the steps and approximations that bridge the gap
between the Continuum Mechanics formulation of Eq.~\ref{eq:Continuum}
and the following FE problem, 

\begin{equation}
\underset{\text{inertial force}}{\underbrace{\boldsymbol{\mathcal{M}}\left(\begin{array}{c}
\ddot{u}_{x}^{{\bf (N-1)}}\\
\ddot{u}_{y}^{{\bf (N-1)}}\\
\vdots\\
\ddot{u}_{x}^{\mathbf{(0)}}\\
\ddot{u}_{y}^{\mathbf{(0)}}
\end{array}\right)}}=\underset{\text{elasticity}}{\underbrace{\boldsymbol{\mathcal{K}}\left(\begin{array}{c}
u_{x}^{{\bf (N-1)}}\\
u_{y}^{{\bf (N-1)}}\\
\vdots\\
u_{x}^{\mathbf{(0)}}\\
u_{y}^{\mathbf{(0)}}
\end{array}\right)}}+\underset{\text{viscosity}}{\underbrace{\boldsymbol{\mathcal{H}}\left(\begin{array}{c}
\dot{u}_{x}^{{\bf (N-1)}}\\
\dot{u}_{y}^{{\bf (N-1)}}\\
\vdots\\
\dot{u}_{x}^{\mathbf{(0)}}\\
\dot{u}_{y}^{\mathbf{(0)}}
\end{array}\right)}},\label{eq:Discrete}
\end{equation}
where the $u_{x}^{\mathbf{(i)}}$'s and $u_{y}^{\mathbf{(i)}}$'s
are the displacements at the nodes $i\in\left\{ 0,\ldots,N-1\right\} $
of a regular mesh.

\subsection{Discretisation of the dynamics\label{app:discretisation_dynamics}}

A central difference scheme is used to discretise Eq.~\ref{eq:Discrete}
in time, \emph{viz.},
\begin{eqnarray}
\delta\dot{u}\left(t_{n}\right) & = & \frac{\delta u\left(t_{n+1}\right)-\delta u\left(t_{n-1}\right)}{2\delta t}+\mathcal{O}(\delta t^{2})\nonumber \\
\delta\ddot{u}\left(t_{n}\right) & = & \frac{\delta u\left(t_{n+1}\right)+\delta u\left(t_{n-1}\right)-2\delta u\left(t_{n}\right)}{\delta t^{2}}+\mathcal{O}(\delta t),\label{eq:iterative_scheme_FEM}
\end{eqnarray}
where $t_{n-1}$, $t_{n}$, and $t_{n+1}$ refer to consecutive time
, separated by a fixed time step $\delta t$.

After insertion into Eq.~\ref{eq:Discrete}, provided that $\delta u(t_{n-1})$
and $\delta u(t_{n})$ are known, the displacements at the next time
step $\delta u(t_{n+1})$ are straightforwardly obtained by inverting
a matrix. The advantage of using a static meshgrid is that this matrix
is then constant and, accordingly, can be inverted once and for all
at the beginning of the simulation.

\subsection{Biperiodic boundary conditions}

We implement biperiodic boundary conditions by connecting the leftmost
nodes of the system to the rightmost ones (see Fig.~\ref{fig:mesh_FEM}),
and the top row to the bottom one.

\section{Relation between the intrinsic macroscopic viscosity and the microscopic
damping coefficient\label{app:viscosity_fitting}}

In MD, the damping magnitude is set by the coefficient $\zeta$ in
the expression of the dissipative force $\boldsymbol{f_{i}}^{D}$
(Eq.~\ref{eq:f_diss_DPD}), whereas it is set by the viscosity $\eta$
in FE. In order to match the damping in both simulations, we must
connect the MD dissipative force $\boldsymbol{f_{i}}^{D}$ to the
viscous stress in FE, namely, $\boldsymbol{\sigma^{\mathrm{diss}}}=2\eta\boldsymbol{\dot{\epsilon}}$
(see Eq.~\ref{eq:Continuum}).

To this end, we consider a pure shear situation, in which particles
are strictly advected by the flow 
\begin{eqnarray*}
\boldsymbol{v}(\boldsymbol{r}) & = & \boldsymbol{\dot{\epsilon}}\cdot\boldsymbol{r}\\
\text{with }\boldsymbol{\dot{\epsilon}} & \equiv & \dot{\epsilon}_{xy}\left(\boldsymbol{e_{y}}\otimes\boldsymbol{e_{x}}+\boldsymbol{e_{x}}\otimes\boldsymbol{e_{y}}\right).
\end{eqnarray*}

\bigskip{}

On the one hand, in MD, the microscopic dissipative stress on particle
$i$ (of volume $V_{0}$) is obtained with the help of the Irving-Kirkwood
formula, \emph{viz.,} 
\begin{eqnarray*}
\boldsymbol{\sigma}(\boldsymbol{r_{i}}) & = & V_{0}^{-1}\sum_{j}\boldsymbol{r_{ij}}\otimes\boldsymbol{f_{ij}^{D}}\\
 & = & -\zeta V_{0}^{-1}\sum_{j}w^{2}\left(r_{ij}\right)\frac{\boldsymbol{v_{ij}}\cdot\boldsymbol{r_{ij}}}{r_{ij}^{2}}\boldsymbol{r_{ij}}\otimes\boldsymbol{r_{ij}}.
\end{eqnarray*}
Focusing on the xy-component of the stress and setting $\boldsymbol{r_{i}}$
as the origin of the frame, \emph{i.e., $\boldsymbol{r_{i}}={\bf 0}$,}
for convenience, we get
\begin{eqnarray}
\sigma_{xy}(\boldsymbol{r_{i}}={\bf 0}) & = & \zeta V_{0}^{-1}\sum_{j}w^{2}\left(r_{j}\right)\frac{\boldsymbol{v_{j}}\cdot\boldsymbol{r_{j}}}{r_{j}^{2}}x_{j}y_{j}\nonumber \\
 & = & \zeta\dot{\epsilon}_{xy}V_{0}^{-1}\sum_{j}w^{2}\left(r_{j}\right)\frac{2y_{j}x_{j}}{r_{j}^{2}}x_{j}y_{j}\nonumber \\
 & \simeq & 2\zeta\dot{\epsilon}_{xy}V_{0}^{-1}\iint ng(r)w^{2}\left(r\right)\frac{x^{2}y^{2}}{r^{2}}d^{2}r\nonumber \\
 & = & 2\zeta n\dot{\epsilon}_{xy}V_{0}^{-1}\int_{0}^{2\pi}\cos^{2}(\theta)\sin^{2}(\theta)d\theta\int_{0}^{\infty}g(r)w^{2}\left(r\right)r^{3}dr\label{eq:sigma_DPD}\\
 & = & \frac{\pi}{2}\zeta n\dot{\epsilon}_{xy}V_{0}^{-1}\int_{0}^{\infty}g(r)w^{2}\left(r\right)r^{3}dr.\nonumber 
\end{eqnarray}
Here, $n$ is the average number density of the system and $g(r)$
is the (alledgedly isotropic) pair correlation function. Equation~\ref{eq:sigma_DPD}
expresses the stress in a volume of space occupied by a particle;
elsewhere the stress is zero. Therefore, the average stress in the
material reads
\begin{eqnarray*}
\overline{\sigma_{xy}} & = & \left(nV_{0}\right)\sigma_{xy}(\boldsymbol{r_{i}}={\bf 0})\\
 & = & \frac{\pi}{2}\zeta\dot{\epsilon}_{xy}n^{2}\int_{0}^{\infty}g(r)w^{2}\left(r\right)r^{3}dr
\end{eqnarray*}

\bigskip{}

On the other hand, in FE, the shear stress simply obeys $\overline{\sigma_{xy}}=2\eta\dot{\epsilon}_{xy}$. 

It immediately follows that

\begin{center}
\framebox{\begin{minipage}[t]{7cm}%
\begin{center}
\begin{equation}
\eta=\frac{\pi}{4}\zeta n^{2}\int_{0}^{\infty}g(r)w^{2}\left(r\right)r^{3}dr.\label{eq:eta_Mono}
\end{equation}

\par\end{center}%
\end{minipage}}
\par\end{center}

If $w^{2}$ decreases fast (but smoothly) and the particles are hard
and dense enough, so that $g\left(r\right)$ exhibits a sharp peak
at $r=a_{0}$, the viscosity in Eq.~\ref{eq:eta_Mono} can be further
approximated as
\begin{eqnarray*}
\eta & \simeq & \frac{1}{8}\zeta n\left(2\pi n\right)\int_{a_{0}-\epsilon}^{a_{0}+\epsilon}g(r)w^{2}\left(r\right)r^{3}dr.\\
 & \simeq & \frac{\zeta nw^{2}\left(a_{0}\right)}{8}\left(2\pi n\right)\int_{a_{0}-\epsilon}^{a_{0}+\epsilon}g(r)r^{3}dr\\
 & \simeq & \frac{1}{8}\zeta nw^{2}\left(a_{0}\right)a_{0}^{2}z_{c},
\end{eqnarray*}
where $z_{c}$ is the coordination number, \emph{i.e.}, the number
of first neighbours (at a distance $r\sim a_{0}$).

\bigskip{}

Equation~\ref{eq:eta_Mono} is valid for a one-component system,
but the extension to binary mixtures, of components A and B, is straightforward;
with transparent notations, the viscosity reads

\begin{center}
\framebox{\begin{minipage}[t]{13cm}%
\begin{center}
\begin{equation}
\eta=\frac{\pi}{4}\zeta\int_{0}^{\infty}\left[n_{A}^{2}g_{AA}(r)+2n_{A}n_{B}g_{AB}(r)+n_{B}^{2}g_{BB}(r)\right]w^{2}\left(r\right)r^{3}dr.\label{eq:eta_Binary}
\end{equation}

\par\end{center}%
\end{minipage}}
\par\end{center}

\selectlanguage{american}%
In the considered Lennard-Jones system, this leads to $\eta=0.726\,\zeta$.

\section{Determination of the local stiffness tensors\label{app:local_stiffness_tensors}}

With our condensed notations for the stress and strain tensors (Eq.~\ref{eq:C3x3_FEM}),
the macroscopic stiffness tensor of an isotropic material of bulk
modulus $K$ and shear modulus $\mu$ reads
\[
{\bf C}=\left(\begin{array}{ccc}
K+\mu & K-\mu & 0\\
K-\mu & K+\mu & 0\\
0 & 0 & 2\mu
\end{array}\right).
\]

\selectlanguage{english}%
In comparison, local stiffness tensors display rather unusual properties.
To grasp the meaning of their (lack of) symmetries, some brief general
considerations about elasticity and deformation are in order.

Suppose that a small macroscopic strain $\bar{\boldsymbol{\epsilon}}$
is applied to a sample and focus on a mesoscopic region $\mathcal{S}$.
The local linear strain tensor $\boldsymbol{\epsilon}$ is defined
as the \emph{symmetric} tensor that best matches the displacements
of the particles in $\mathcal{S}$ due to the applied strain. Only
if the deformation is strictly affine over the whole sample do the
local strain tensors equate to $\bar{\boldsymbol{\epsilon}}$.

Because, for a given short-range interparticle potential, the local
stress $\boldsymbol{\sigma}$ results from the local configuration
of particles, it is reasonable (but not strictly necessary) to suppose
the existence of a function $f$ such that
\[
\boldsymbol{\sigma}=f\left(\boldsymbol{\epsilon}\right).
\]
Let us write the first-order Taylor expansion of $f$, provided that
it exists,

\begin{equation}
\sigma_{\alpha\beta}-\sigma_{\alpha\beta}^{(0)}=C_{\alpha\beta\gamma\delta}\epsilon_{\gamma\delta}+\mathcal{O}\left(\left\Vert \boldsymbol{\epsilon}\right\Vert ^{2}\right),\label{eq:Stiffness1}
\end{equation}
where $\alpha,\,\beta\in\left\{ x,y\right\} $ and $\sigma_{\alpha\beta}^{(0)}$
is the quenched stress in the original configuration. With condensed
notations, Eq.~\ref{eq:Stiffness1} turns into%
\footnote{As a minor technical detail, note that, because the tensorial multiplication
$C_{\alpha\beta\gamma\delta}\epsilon_{\gamma\delta}$ involves a summation
on both $\epsilon_{xy}$ and $\epsilon_{yx}$, components $C_{\alpha\beta,\gamma\delta}$
of the \emph{second-rank} tensor ${\bf C}$ may not exactly equate
to their counterparts in the \emph{fourth-rank} tensor $C_{\alpha\beta\gamma\delta}$;
for instance, $C_{xy,xy}=2C_{xyxy}$.%
}

\begin{equation}
\left(\begin{array}{c}
\sigma_{xx}\\
\sigma_{yy}\\
\sqrt{2}\sigma_{xy}
\end{array}\right)=\underset{{\bf C}}{\underbrace{\left(\begin{array}{ccc}
C_{xx,xx} & C_{xx,yy} & C_{xx,xy}\\
C_{yy,xx} & C_{yy,yy} & C_{yy,xy}\\
C_{xy,xx} & C_{xy,yy} & C_{xy,xy}
\end{array}\right)}}\left(\begin{array}{c}
\epsilon_{xx}\\
\epsilon_{yy}\\
\sqrt{2}\epsilon_{xy}
\end{array}\right)+\mathcal{O}\left(\left\Vert \boldsymbol{\epsilon}\right\Vert ^{2}\right).\label{eq:Stiffness2}
\end{equation}
The affine strain-local stress approximation consists in replacing
the components of $\boldsymbol{\epsilon}$ on the rhs of Eq.~\ref{eq:Stiffness2}
with those of the affine strain $\bar{\boldsymbol{\epsilon}}$, in
order to determine ${\bf C}$ more easily. For subregions of size
larger than $5\sigma_{AA}$, \citet{Mizuno2013moduli} showed that
this approximation is quite reasonable, although it slightly underestimates
the spatial fluctuations of the elastic constants. On the other hand,
should the local stress on the lhs be computed for a \emph{local}
deformation equal to $\boldsymbol{\bar{\epsilon}}$, \emph{i.e.},
should the system not be allowed to relax to the energy minimum after
the application of the affine strain $\boldsymbol{\bar{\epsilon}}$,
then we would obtain the so-called Born term ${\bf C}^{B}$, which
largely overestimates the stiffness of the disordered material \citep{Mizuno2013moduli}.

\selectlanguage{american}%
\medskip{}

\begin{figure}
\begin{centering}
\subfloat[\selectlanguage{american}%
Asymmetry.\selectlanguage{english}
]{\includegraphics[width=7cm]{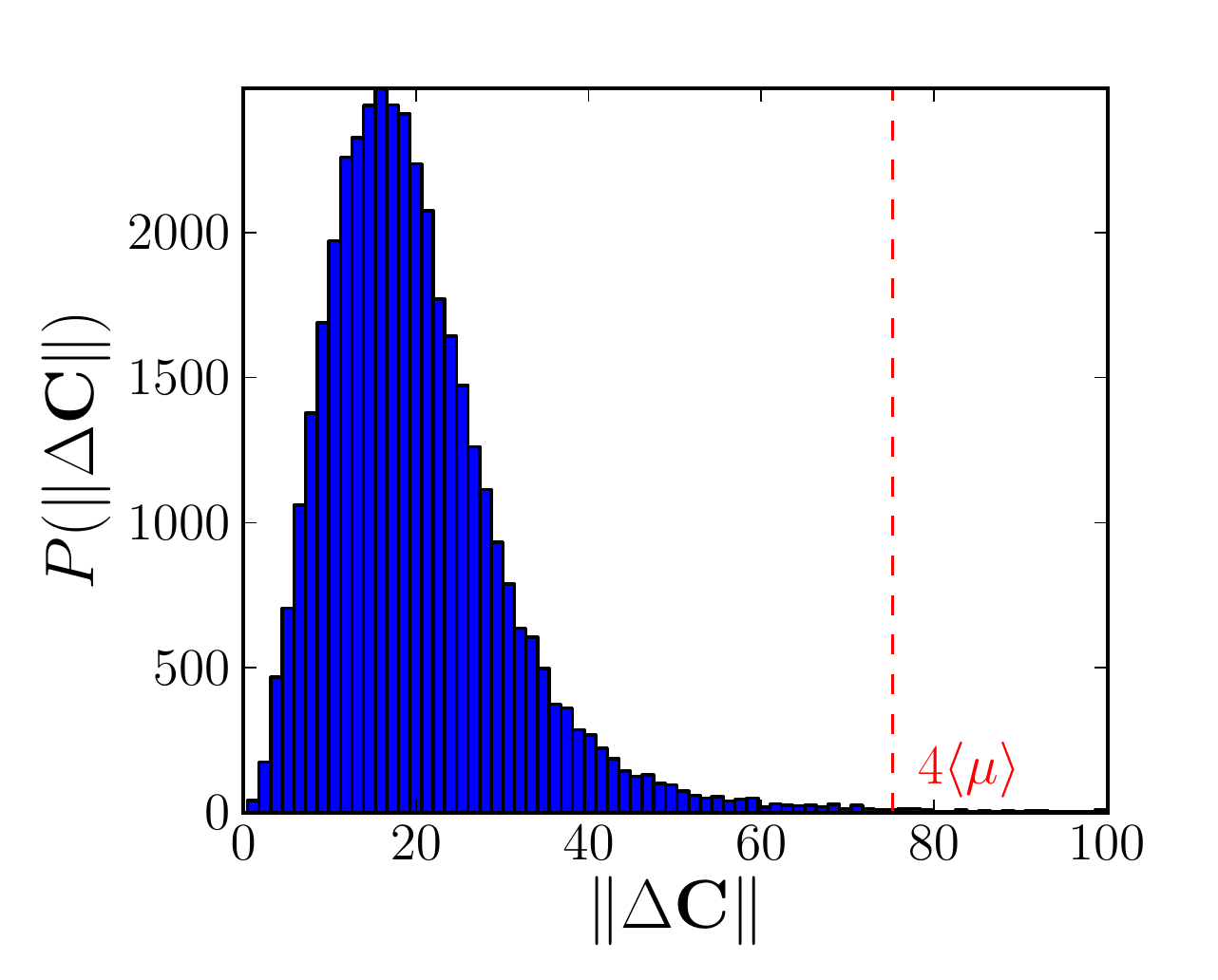}

}\subfloat[\selectlanguage{american}%
Discrepancy with the matrix ${\bf C}^{\prime}$ given in Eq.~\ref{eq:Cprime_FEM}.\selectlanguage{english}
]{\includegraphics[width=7cm]{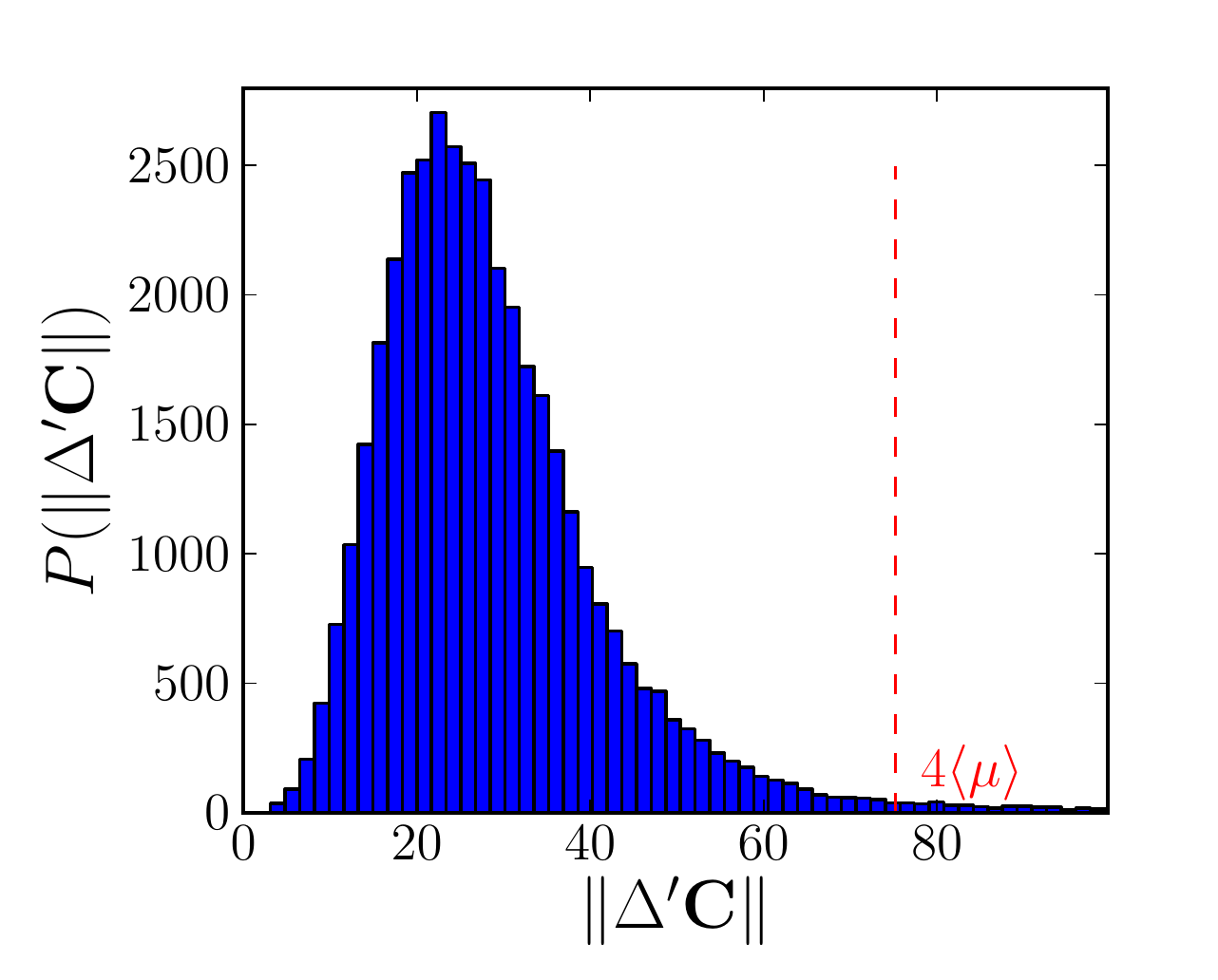}

}
\par\end{centering}

\caption{\label{fig:Dev_due_to_C_proj}Histograms of the approximation errors
made when supposing that the local stiffness tensors ${\bf C}$ are
(a) symmetric, (b) of the form given in \foreignlanguage{american}{Eq.~\ref{eq:Cprime_FEM}.}}
\end{figure}

\selectlanguage{english}%
For the time being, all components of the second-rank stiffness tensor
${\bf C}$ are independent. But, if the local stress derives from
a (twice differentiable) local strain-energy density $e$, \emph{i.e.},
\[
\sigma_{\alpha\beta}\equiv\frac{\partial e}{\partial\epsilon_{\alpha\beta}},
\]
then 
\[
C_{\alpha\beta\gamma\delta}=\frac{\partial^{2}e}{\partial\epsilon_{\alpha\beta}\partial\epsilon_{\gamma\delta}}.
\]
It immediately follows that $C_{\alpha\beta\gamma\delta}=C_{\gamma\delta\alpha\beta}$;
this symmetry property is transferred to the second-rank tensor ${\bf C}$
(thanks to the carefully chosen $\sqrt{2}$ prefactors in Eq.~\ref{eq:Stiffness2}).
Indeed, \citet{Tsamados2009} observed numerically that, for coarse-graining
regions larger than 5 Lennard-Jones particles in diameter, assuming
a symmetric stiffness matrix ${\bf C}$ creates an error of less than
1\% on the local stress evaluations. In the MD system under consideration,
we quantify the asymmetry of the mesoscopic stiffness matrices, computed
over regions of size $a=5\sigma_{AA}$, with the following measure:

\[
\left\Vert \Delta{\bf C}\right\Vert \equiv\sqrt{\underset{^{\left\{ xx,yy,xy\right\} }}{\sum_{i,\, j\,\in}}\Delta C_{i,j}^{2}}\text{ with }\Delta{\bf C}\equiv{\bf C}-\frac{{\bf C}+{\bf C}^{\top}}{2}.
\]
What should $\left\Vert \Delta{\bf C}\right\Vert $ be compared with?
At first sight, the answer would be $\left\Vert {\bf C}\right\Vert $,
but the latter is dominated by large symmetric terms involving the
bulk modulus $K\approx100$. Thus, on second thoughts, it appears
more informative to remove the terms involving $K$; $\left\Vert \Delta{\bf C}\right\Vert $
should then be compared to, \emph{e.g.}, $\left\langle \mathrm{Tr}\left({\bf C}\right)-2K\right\rangle =4\left\langle \mu\right\rangle $,
with $\left\langle \mu\right\rangle =18.8$. From the histogram of
$\left\Vert \Delta{\bf C}\right\Vert $ values plotted in Fig.~\ref{fig:Dev_due_to_C_proj}a,
it transpires that deviations from symmetry in ${\bf C}$ are not
strictly negligible, but symmetry may nevertheless be a decent \emph{approximation}.

\selectlanguage{american}%
\medskip{}

\selectlanguage{english}%
To further reduce the number of local parameters, the isotropic contraction/dilation
vector $(\nicefrac{\sqrt{2}}{2}\,\nicefrac{\sqrt{2}}{2}\,0)^{\top}$
is supposed to produce an isotropic compression and, thus, to be an
eigenvector of ${\bf C}$, \emph{ergo}
\[
\begin{cases}
C_{xy,xx} & =-C_{xy,yy}\\
C_{xx,xx} & =C_{yy,yy}
\end{cases}
\]
The assumptions of tensorial symmetry and isotropic response to contraction
come down to projecting ${\bf C}$ onto a matrix of the form
\begin{equation}
{\bf C}^{\prime}=\left(\begin{array}{ccc}
\alpha & \delta & \beta\\
\delta & \alpha & -\beta\\
\beta & -\beta & \upsilon
\end{array}\right)\text{ with }\alpha,\delta,\beta,\upsilon\in\mathbb{R},\label{eq:Cprime_FEM}
\end{equation}
where $\alpha$ and $\beta$ will be the averages of the pairs $\left(C_{xx,xx},C_{yy,yy}\right)$
and $\left(C_{xy,xx},-C_{xy,yy}\right)$, respectively. The approximation
error, quantified by $\left\Vert \Delta^{\prime}{\bf C}\right\Vert \equiv\left\Vert {\bf C}-{\bf C}^{\prime}\right\Vert $,
is plotted in Fig.~\ref{fig:Dev_due_to_C_proj}b. As expected, the
deviations are somewhat larger than were ${\bf C}$ only symmetrised,
but they remain under control.

\selectlanguage{american}%
\medskip{}

\selectlanguage{english}%
For each matrix ${\bf C}^{\prime}$, we compute the eigenvalues $c_{1}\leqslant c_{2}\leqslant c_{3}$
and define:

- the small local shear modulus $\mu_{1}\equiv c_{1}/2$, 

- the large local shear modulus $\mu_{2}\equiv c_{2}/2$, 

- and the bulk modulus is $K\equiv c_{3}/2$. 

\begin{figure}
\begin{centering}
\subfloat[$\mu_{1}$]{\includegraphics[width=4cm]{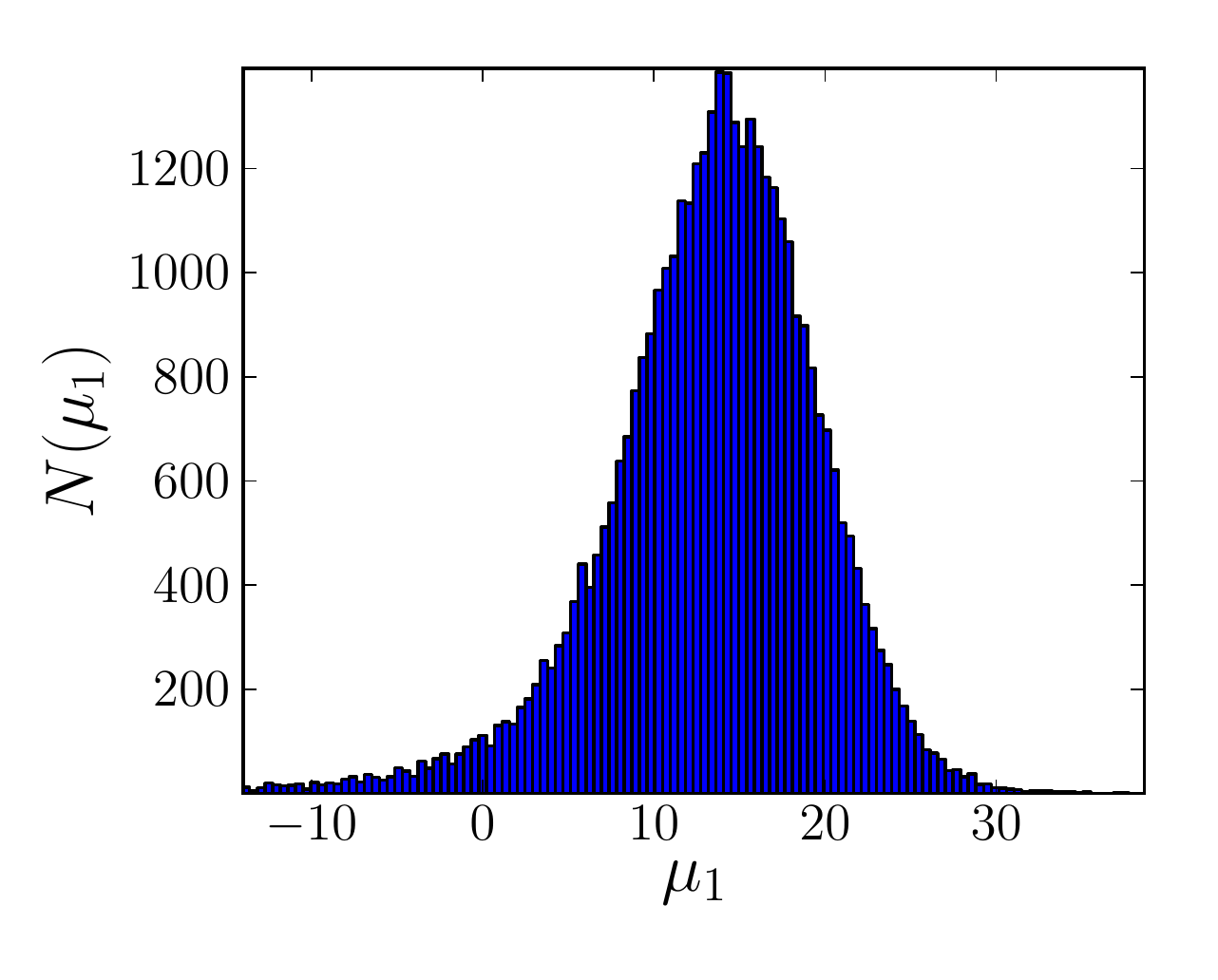}

}\subfloat[$\mu_{2}$]{\includegraphics[width=4cm]{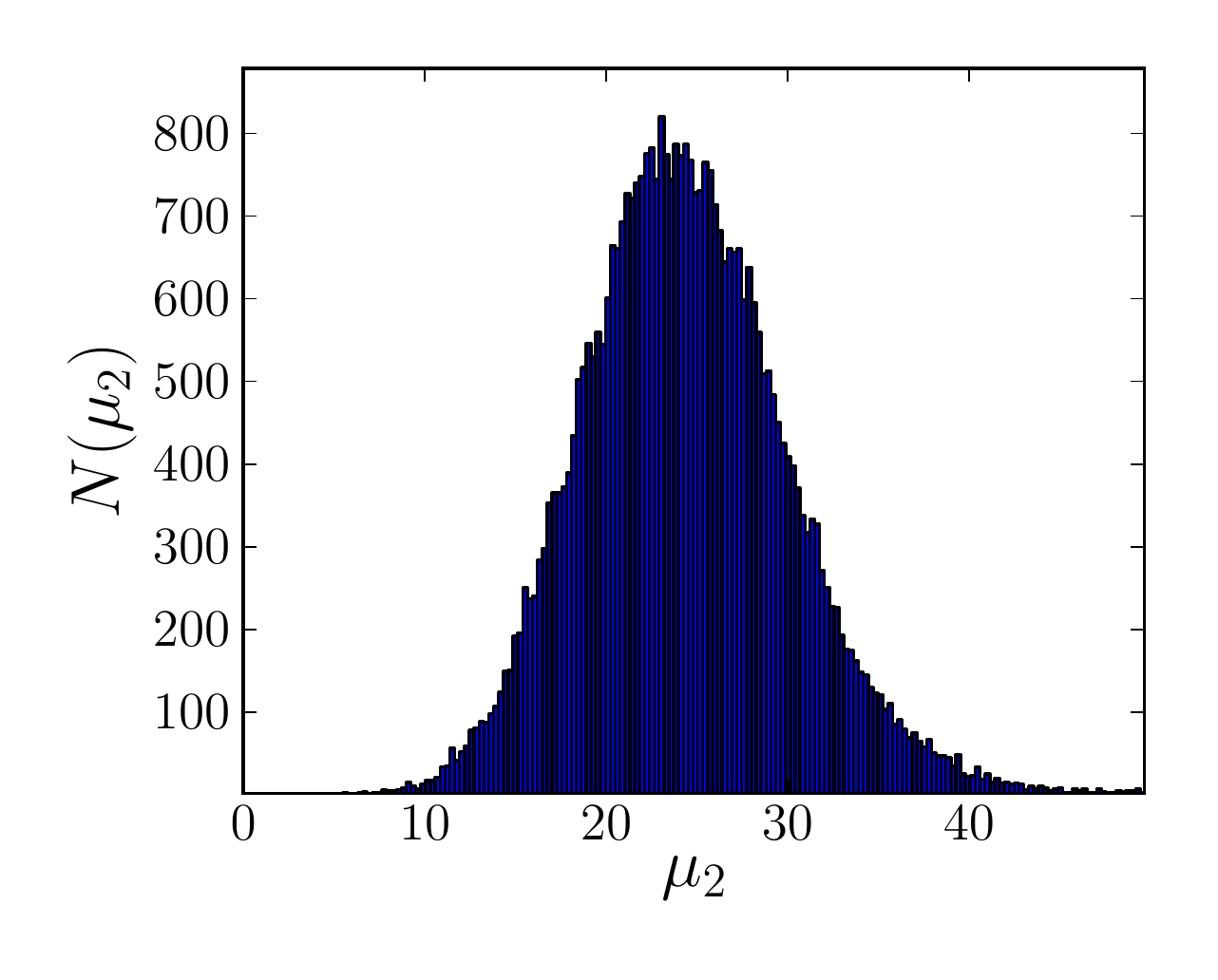}

}\subfloat[$K$]{\includegraphics[width=4cm]{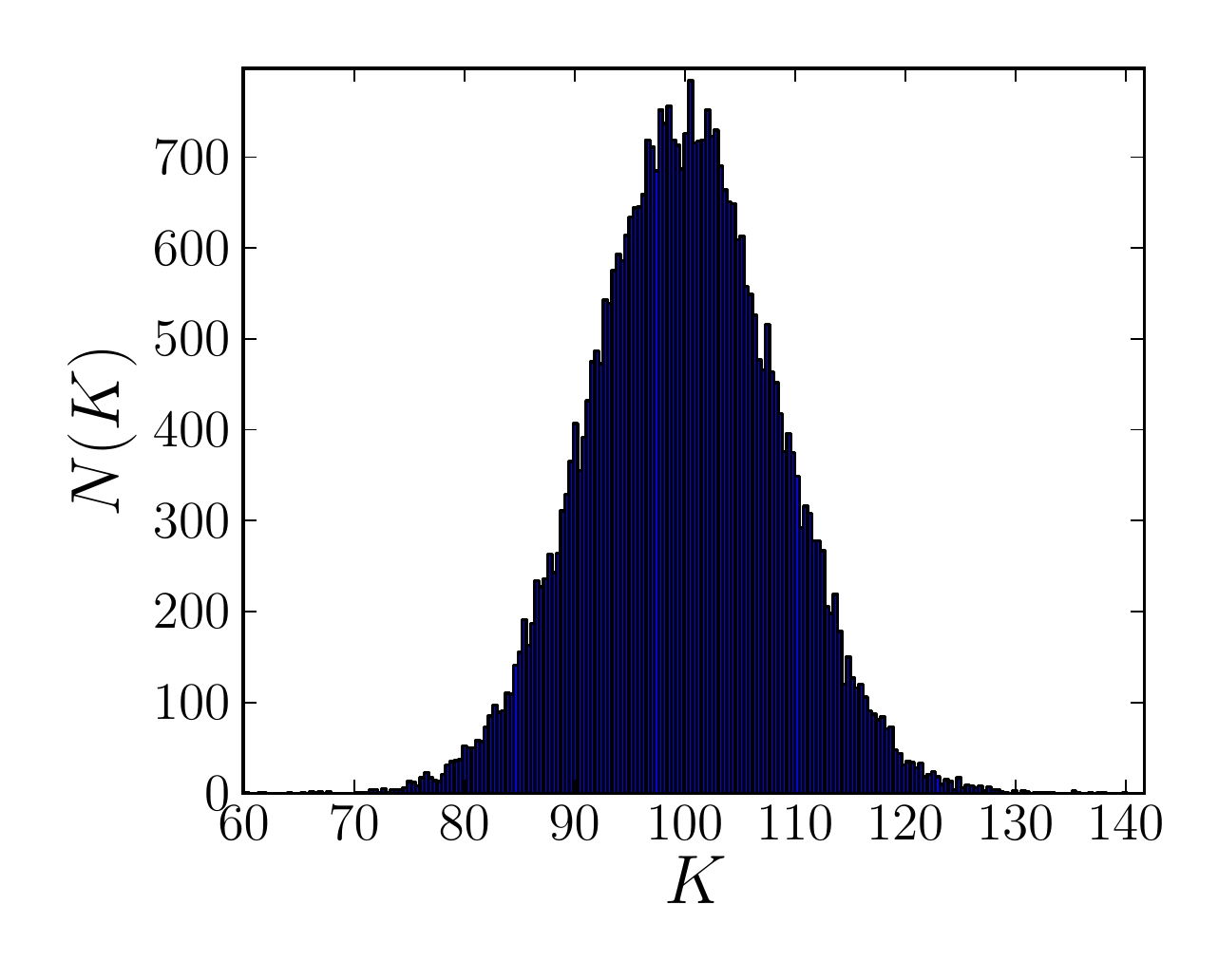}

}
\par\end{centering}

\caption{\label{fig:Dist_local_el_csts}Histograms (number of counts) of the
measured values of the local elastic constants $\mu_{1}$, $\mu_{2}$,
and $K$ in subregions of size $5\sigma_{AA}\times5\sigma_{AA}$ in
the MD system.}
\end{figure}

The distributions of these local elastic constants are presented in
Fig.~\ref{fig:Dist_local_el_csts} and their mean values and standard
deviations are summarised in Table~\ref{tab:Statistical-prop}. It
should be noted that the average eigenvalues of the projected tensor
${\bf C^{\prime}}$ differ by $10\%$ or less from the eigenvalues
of the full local stiffness tensors ${\bf C}$. 

The components of ${\bf C^{\prime}}$ can then be rewritten as follows

\selectlanguage{french}%
\begin{eqnarray*}
\begin{cases}
\alpha & \equiv K+\mu_{2}\cos^{2}2\theta+\mu_{1}\sin^{2}2\theta\\
\delta & \equiv K-\mu_{2}\cos^{2}2\theta-\mu_{1}\sin^{2}2\theta\\
\beta & \equiv\frac{\sin4\theta}{\sqrt{2}}\left(\mu_{2}-\mu_{1}\right)\\
\upsilon & \equiv2\mu_{2}\sin^{2}2\theta+2\mu_{1}\cos^{2}2\theta
\end{cases} & ,
\end{eqnarray*}
where the angle $\theta$ has been defined in Section~\ref{sec:local_el_csts}.

\selectlanguage{english}%

\end{document}